\documentclass{aa}  

\usepackage{graphicx}
\usepackage{txfonts}
\usepackage{longtable}
\usepackage{subcaption}
\usepackage{float}
\usepackage{adjustbox}

\usepackage{geometry}
\usepackage{supertabular}                                             
\usepackage[table]{xcolor}
\usepackage{overpic}

\usepackage{orcidlink}          
\usepackage{hyperref}
\hypersetup{
    colorlinks=true,
    linkcolor=blue,
    citecolor=blue,
    urlcolor=blue
}

\usepackage{multirow}

\usepackage{color}
\definecolor{red(new)}{RGB}{200, 50, 60}

\begin{document}

   \title{Ensemble pulsational characteristics of eclipsing binaries containing $\beta$ Cep stars observed by the TESS mission}

   \titlerunning{Ensemble pulsational characteristics of eclipsing binaries containing $\beta$ Cep stars observed by the TESS mission}

   \author{C. I. Eze \orcidlink{0000-0003-3119-0399} \inst{\ref{CAMK},\ref{UNN},\ref{KUL}} \fnmsep\thanks{Corresponding author} 
        \and
        G. Handler \orcidlink{0000-0001-7756-1568} \inst{\ref{CAMK}} 
        \and
        M. Vanrespaille \orcidlink{0000-0002-0420-2473}
         \inst{\ref{KUL}}
         \and
       A. Kemp \orcidlink{0000-0003-2059-5841} \inst{\ref{KUL}}
          \and
        A. Tkachenko \orcidlink{0000-0003-0842-2374} \inst{\ref{KUL}}
        \and
        D. M. Bowman \orcidlink{0000-0001-7402-3852} \inst{\ref{NCL},\ref{KUL}}
        \and\\
        C. Aerts \orcidlink{0000-0003-1822-7126} \inst{\ref{KUL},\ref{Rad},\ref{MPIA}}        }

   \authorrunning{C.I. Eze et al.}
   \institute{Nicolaus Copernicus Astronomical Centre, Polish Academy of Sciences, Bartycka 18, PL-00-716 Warsaw, Poland \\
           \email{cheze@camk.edu.pl} \label{CAMK} \and
           Department of Physics and Astronomy, Faculty of Physical Sciences, University of Nigeria, Nsukka, Nigeria\label{UNN} 
    \and
     Institute of Astronomy, KU Leuven, Celestijnenlaan 200D, 3001 Leuven, Belgium \label{KUL}
    \and School of Mathematics, Statistics and Physics, Newcastle University, Newcastle upon Tyne, NE1 7RU, UK \label{NCL}
    \and
Department of Astrophysics, IMAPP, Radboud University Nijmegen, PO Box 9010, 6500 GL Nijmegen, The Netherlands \label{Rad} 
         \and
              Max Planck Institut f\"ur Astronomie, K\"onigstuhl 17, 69117 Heidelberg, Germany \label{MPIA}       
    }



   \abstract
   {Massive pulsators in eclipsing binaries provide precise, model-independent stellar parameters that, combined with asteroseismic constraints, provide powerful probes of internal rotation, convective-core mass, angular-momentum transport, and mixing in massive stars.}
   {This study investigates the pulsational characteristics and their impact on the structure and evolution of $\beta$ Cep pulsators in eclipsing binaries.}
   {We analyse an ensemble of 65 $\beta$ Cep stars in eclipsing binaries, identify rotationally split modes where possible, and statistically derive the dependence of pulsational properties on binary orbital parameters. The interior properties of the systems were also derived using pre-computed grids of stellar structure models with \textsc{MESA} and associated pulsation frequencies using \textsc{StORM} and an asteroseismic routine that tracks the age and rotation of the star in the entire $\beta$ Cep instability region.}
   {Interior rotation rates are derived for 14 systems that exhibit rotational mode splitting. While differential rotation is tentatively inferred for HD~112485, quasi-rigid rotation is found for the primaries of other systems except CZ~Vel. Detailed forward asteroseismic modelling was conducted for HD~112485 and EK~Cru to deduce their seismic and core properties, as well as their evolutionary state. HD 112485 is one of the least evolved $\beta$ Cep stars asteroseismically modelled to date, with a core hydrogen mass fraction $X_{c}= 0.501_{-0.013}^{+0.009}$. The two reproduced zonal modes in HD~112485 are the first radial-order dipole pressure mode and the second radial-order quadrupole gravity mode, while the zonal modes in EK Cru are the second-order quadrupole mixed mode and the third order dipole pressure mode. A multivariate regression shows a negative correlation of the pulsation amplitude with pulsation frequency, and no statistically significant relationship with the orbital period. In addition, no significant relationship is observed between the pulsation frequency and the orbital period in the current sample.}
   {}



   \maketitle
%

\section{Introduction}

  Massive stars are predominantly found in multiple systems \citep{Sanaetal2012, Sanaetal2014, Kobulnickyetal2014, Offner2023}. Hence, a substantial fraction ($\geq 60\%)$ are expected to be in binaries with over 17\% reported to be in eclipsing binaries \citep{Chinietal2013}. Also, a substantial fraction of the massive eclipsing binaries shows component(s) that are pulsating, which allow us to probe the physics of stellar structure and evolution via combined strengths of eclipsing binary modelling and asteroseismology --- see review by \citet{SouthworthandBowman2026}. These pulsations manifest in the form of pressure modes (p modes) and gravity modes (g modes) \citep{Unnoetal1989}, mixed modes \citep{Dupretetal2009}, Rossby modes (r modes; e.g. \citealt{PapaloizouandPringle1978, GittinsandAndersson2023}), strange modes \citep[e.g.][]{Papaloizouetal1997, Saioetal1998}, and overstable convective modes \citep{Lee2021} and gravito-inertial modes in fast rotators \citep{AertsandTkachenko2024}. One of the subclasses of massive pulsators that shows some of these pulsations is the $\beta$ Cep stars class. These pulsators have masses ranging from about 6 to upwards of 30~M$_{\odot}$ \citep{Burssensetal2020, Fritzewskietal2025}. These pulsators show predominantly low-radial order p and g modes. This makes them excellent for massive star asteroseismology, enabling one to probe the entire star \citep[see][]{StankovandHandler2005, Bowman2020}.   

  Eclipsing binaries are particularly powerful laboratories for testing stellar structure and evolution theory, since the application of Kepler's laws of motion to their combined light curve and radial velocity time series yields model-independent masses and radii (see \citealt{Southworth2021}). The absolute numbers of massive pulsators reported in binary or multiple eclipsing systems are significantly smaller compared to their low-mass counterparts \citep{Kirketal2016, Ijspeertetal2021, Prsaetal2022}. However, since the advent of the Transiting Exoplanet Survey Satellite (TESS) mission \citep{Rickeretal2015}, there has been a concerted effort to increase the sample size of massive pulsators in eclipsing binaries and conduct ensemble asteroseismology to improve the physics of stellar structure and evolution models. This major barrier of small sample size thus far has hampered the general and ensemble asteroseismic understanding of massive stars. Although not completely solved, the TESS mission has provided the largest to-date catalogue of such stars: 78 $\beta$ Cep pulsators in eclipsing binaries, among which are 59 new discoveries \citep[e.g.][]{EzeandHandler2024b,EzeandHandler2024a}. These eclipsing binaries containing a $\beta$~Cep star span different morphological classifications ranging from possible contact to detached binaries, 
  allowing to deduce the effect of binary dynamics on pulsation properties in different scenarios. These can also be used to test different scenarios of binary evolution (e.g. pre- and post-interaction).

 To date, effectively all asteroseismic analyses of $\beta$ Cep stars have assumed single-star evolution theory when interpreting the observed pulsation frequencies \citep[e.g.][]{Aertsetal2003, Handleretal2004, Burssensetal2023, Vanlaer2025}. Beyond studies of individual $\beta$ Cep stars, \citet{Fritzewskietal2025} conducted ensemble asteroseismic forward modelling of 119 $\beta$ Cep stars using a combination of $\it{TESS}$ and $\it{Gaia}$ light curve data, but the sample only contained three previously known eclipsing binaries. Considering the possible effect of binary dynamics (e.g. orbital angular momentum transport and/or mass transfer) on the pulsational properties of stars, it is not yet clear whether the findings of \citet{Fritzewskietal2025} properly account for the pulsational properties of $\beta$ Cep pulsators in eclipsing binaries. Examples of such effects of binarity on the pulsations of $\beta$~Cep stars include pulsation amplitude dilution of $\beta$ Cep components in eclipsing binaries \citep{Handleretal2005}, and tidal suppression of pulsations in close binary systems, such as in the case of $\delta$~Sct stars in eclipsing binaries \citet{LiakosandNiarchos2017}. However, preliminary results do not show pulsation--orbital periods relation(s) in $\beta$ Cep stars in eclipsing binaries \citep{Ezeetal2025} contrary to what is obtainable in $\delta$~Sct stars. Moreover, some $\beta$~Cep stars in close binaries have pulsations that are significantly affected by the equilibrium tide because of a close companion (e.g. \citealt{Southworthetal2020, Southworthetal2021}). All of these studies indicate the need to conduct a detailed ensemble analysis of $\beta$ Cep stars in eclipsing binaries to decipher the pulsational characteristics of this class of stars in binaries.      

In this work, we aim to identify the pulsation modes, where tenable, and investigate the ensemble pulsational characteristics of the sample of $\beta$ Cep pulsators in eclipsing binaries observed by {\it TESS} from \citet{EzeandHandler2024a}. In Section \ref{sec:sample description}, we describe the sample, its observational properties, and selection criteria. We describe the methods used for pulsation frequency extraction in Section \ref{sec:pulsfreq_analysis}. In Section \ref{sec:statistical_characterisation}, we describe the statistical characterisation of the sample. We describe mode identification procedures in Section \ref{sec:mode_identification}, discuss tidal synchronisation of the systems in Section \ref{sec:synchronisation},  discuss the regression analysis of the pulsation and orbital parameters in Section \ref{sec:reg_analysis}, perform asteroseismic modelling of selected cases in Section \ref{sec:asteroseismic_modelling} and conclude in Section \Ref{sec:conclusion}.

\section{Sample Description and Observation}\label{sec:sample description}
The data sources for this analysis are the catalogue of $\beta$ Cep stars in eclipsing binaries published by \citet{EzeandHandler2024b} and the 2-min SPOC light curves from the TESS mission \citep{Rickeretal2015}. In the catalogue, there are 78 reported $\beta$ Cep pulsators in eclipsing binaries. To obtain a more homogeneous sample and to avoid outliers from unduly driving the results, three systems in the catalogue with no clear orbital period or targets with extremely long orbital period or extremely  high pulsation amplitude were removed. In particular, this concerns one star with such a high amplitude (161 mmag) for its pulsation that it would affect the statistical results too heavily if left in the sample. Ten other reported $\beta$ Cep candidates in the catalogue were also excluded in this analysis because their nature could not be explicitly confirmed in the current data. This results in a sample of 65 confirmed $\beta$ Cep pulsators in eclipsing binaries analysed in this work. The available  atmospheric solutions  and the absolute parameters of the systems used in this work were obtained predominantly from Eze et al. (A\&A, submitted) and with a few obtained from the literature \citep[e.g.][]{Cakirlietal2025}. 

The 2-min TESS SPOC light curves used for the analysis were extracted from the MAST portal, normalised, and cleaned of systematic and instrumental noise.
Each sector was independently median-normalised prior to merging. No additional polynomial or spline detrending step was applied, since the subsequent pulsation analysis focused on the range of frequencies $3 -14\, $d$^{-1}$, characteristic of $\beta$ Cep pulsations and far above the low frequency regime,  where residual instrumental or sector-stitching trends are expected to dominate. Only frequencies within the stated range were used for mode identification or rotational-splitting analysis. 
In systems where 2-min light curves are unavailable (8 targets), the TESS QLP data \citep{2020RNAAS...4..204H,2020RNAAS...4..206H} of the targets are used instead.

\section{Pulsation Frequency extraction}\label{sec:pulsfreq_analysis}
We first examined the normalised light curves and removed instrumental noise and outliers.
Independent pulsations were extracted from the cleaned light curves using \textsc{STAR SHADOW} \citep{Ijspeertetal2024} and \textsc{Period04} \citep{LenzandBreger2005}.  Although \textsc{STAR SHADOW}  and \textsc{Period04} are independent pulsation analysis codes, 
we used them complementarily for validation purposes.  For systems in dense fields, with neighbouring stars of comparable magnitudes, where blending is suspected, the accepted  prewhitened frequencies from \textsc{STAR SHADOW} were checked with \textsc{TESS-Localise} \citep{HigginsandBell2022}, which localises the likely origin of the pulsation(s), before their acceptance as independent pulsations from the targets. Possible combination frequencies  were also removed. Finally, significant $\beta$  Cep pulsation frequencies that are common to both  \textsc{STAR SHADOW} and \textsc{Period04}  analyses were accepted as independent pulsations in this work.

\textsc{STAR SHADOW} in its default  configuration considers a frequency significant if the Bayesian Information Criterion (BIC)  decreases by 2 \citep{Ijspeertetal2024, Fritzewskietal2025}. Although it is capable of analysing blended peaks \citep{Ijspeertetal2024}, it could add noise peaks  or many harmonics to the model, resulting in over-extraction of frequencies.  For systems in eclipsing binaries, for which \textsc{STAR SHADOW} is specifically built, this is a benefit as it helps to build more accurately the binary harmonic model and solve the binary orbital dynamics. Although this has the  advantage of building binary eclipse models more precisely, it may introduce spurious pulsation frequencies. To reduce this effect, we extracted significant $\beta$  Cep frequencies using BIC=10 in line with \citet{Fritzewskietal2025}, such that \textsc{STAR SHADOW} considers a frequency significant only when the BIC decreases by at least 10. This reduces the number of prewhitened frequencies significantly without leaving out significant high amplitude pulsations. Frequencies are adjudged significant in line with \citet{VanBeecketal2021}. 
See \citet{Ijspeertetal2024}  and \citet{Fritzewskietal2025} for details of analysis with \textsc{STAR SHADOW}. For our analysis with \textsc{STAR SHADOW}, the orbital periods reported in \citet{EzeandHandler2024b} were used for the analysis to build the harmonic models, which are subsequently subtracted from the light curves to obtain independent pulsations.  The orbital periods of CD-51 9984 and HD 157400 were taken from Eze et al. (submitted).

The pulsation frequencies were also independently prewhitened with Period04, after removing the effect of binarity in the light curves. To this end, a two-fold strategy was adopted. For binaries with orbital periods below 10\,d, a 100-frequency multiharmonic fit based on the orbital frequency was computed first, and all signals with an amplitude exceeding a signal-to-noise ratio (S/N) of 3.5 were kept. This heuristic fit was subtracted from the data and the residuals were searched for statistically significant pulsation frequencies, according to the criterion of \cite{2021AcA....71..113B}. For systems with orbital periods in excess of 10\,d, the eclipses were simply removed from the data and only the out-of-eclipse light curves searched for pulsations adopting the same detection criterion.

The search for $\beta$ Cep frequencies in this analysis was restricted to $3\leq f\leq14$~d$^{-1}$ to avoid possible contamination by SPB and $\delta$~Sct pulsations of companions, thus ensuring that the subsequent analysis was based primarily on $\beta$ Cep pulsation modes. The list of the accepted significant independent $\beta$ Cep pulsations used for the subsequent analysis is shown in Appendix \ref{appendix:tab-rot-multiplet}.


\section{Statistical Characterisation}\label{sec:statistical_characterisation}
\begin{figure*}[htbp]
\centering
\includegraphics[width=0.42\linewidth]{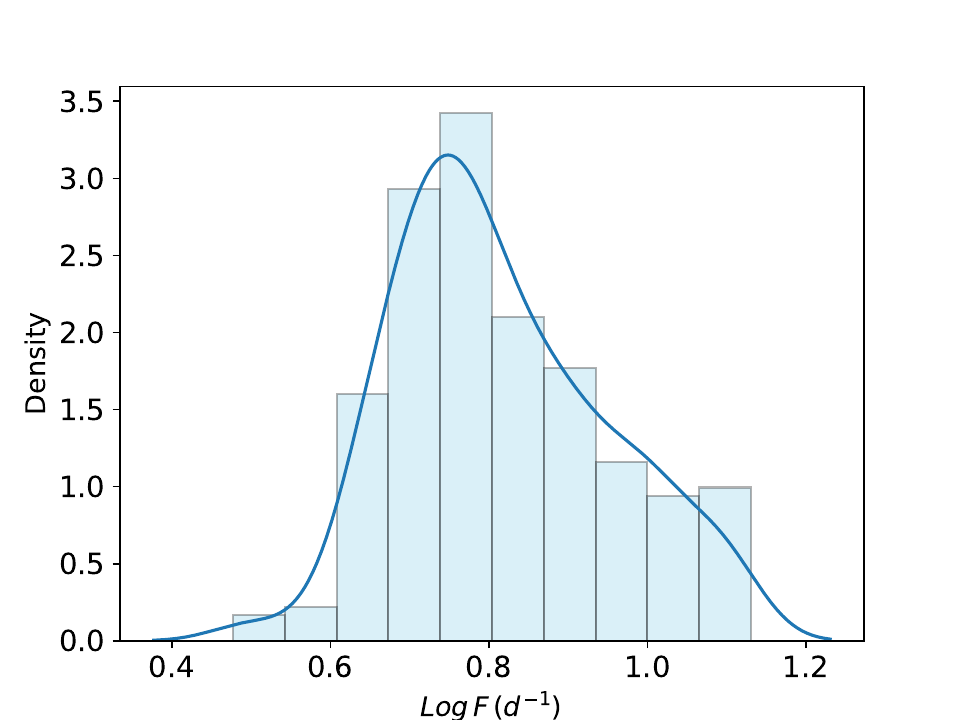}
\includegraphics[width=0.42\linewidth]{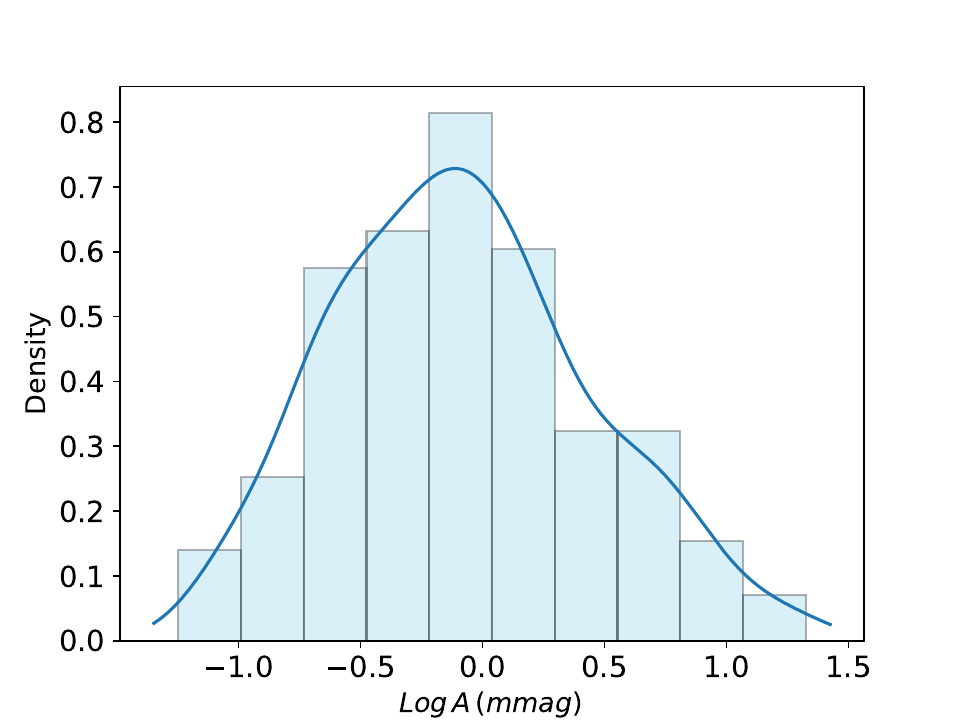}
\caption{Distributions of pulsation frequency (left panel) and pulsation amplitude (right panel) in log10 scale.}\label{fig:fig_kde}

\end{figure*}

We examine the statistical distributions of the observed pulsation frequencies and amplitudes across the ensemble. Figure \ref{fig:fig_kde} shows histograms of the pulsation frequencies (left panel) and amplitudes (right panel) over-plotted  with their Kernel density estimation. This
is a non-parametric method of estimating the probability density function of a random variable \citep{Rosenblatt1956, Parzen1962, XieandWu2014} and helps to visualise the range and distributions of a random variable. It traces a continuous curve that does not depend on the chosen number of bins like a histogram. As evident in the distribution plots, both the pulsation frequencies and the pulsation amplitudes show unimodal distributions. These suggest that the frequencies and amplitudes are each consistent with being drawn from a single underlying population. Since the amplitude spans multiple orders of magnitude, we used $\log_{10}$ scale for the analysis to reduce dispersion.  Both distributions are positively skewed with that of the frequency being more pronounced. 

The pulsation frequencies occur mainly in the range of $0.477 -1.13$ in $\log_{10}(f/\mathrm{d}^{-1})$, corresponding to a range of 
$3 - 13.51$~d$^{-1}$, following our preselection criteria. This range is fully in line with the literature for known apparently single $\beta$ Cep stars (e.g. \citealt{Shietal2024}). The pulsation frequencies show a modal peak at  $\log\,f \approx 0.749$  $(f \approx 5.53$~d$^{-1}$) with a median pulsation frequency, $f \approx 6.21$~d$^{-1}$, resulting in a median pulsation period of $P_{\rm puls}= 1/f \approx 0.16$\,d. This is in fairly good agreement with \citet{StankovandHandler2005} and \citet{PigulskiandPojmanski2009}, who studied ensembles of apparently single $\beta$ Cep stars and found unimodal distributions of the pulsation period with a median pulsation period of approximately $0.17$~d.  However, while they found a symmetric unimodal distribution around the median, the distribution reported in this work for the pulsation frequency appears to be asymmetric around the median and positively skewed (in both $\log$ and nominal scales), thus favouring the observation of lower frequencies compared to higher ones.   Radiative damping or geometric cancellation effects at higher frequencies \citep[e.g.][]{Dziembowski1977, Belkacemetal2012} could possibly explain this skewed  distribution of pulsations.  The departure from normality implies that although  pulsations could be approximated (especially the amplitude) to normal distributions, they do not strictly follow normal distributions.
The amplitudes, on the other hand, have a median value of approximately 0.76 mmag, respectively, consistent with the prevalence of low amplitude pulsations in $\beta$ Cep stars.  

\section{Mode Identification}\label{sec:mode_identification}
The pulsation modes of the eclipsing binary systems were identified through the search for rotational multiplets in TESS light-curve data. 
For $\beta$ Cep non-radial pulsation modes, the observed frequency of a mode with quantum numbers 
$(n, \ell, m)$ is defined, following a first-order perturbative treatment of rotational effects  \citet{Ledoux1951, DziembowskiandGoode1992, Soufi1998} as:

\begin{equation}
f_{n\ell m} =
f_{n\ell,0} + m \left(1 - C_{n\ell} \right)\Omega
+ \Omega^2 \left( D_{n\ell} + m^2 E_{n\ell} \right) ~ ,
\label{eqn:ledoux}
\end{equation}

\noindent
where $f_{n\ell m}$ is the observed oscillation frequency of a mode with radial order $n$, spherical degree $\ell$, and azimuthal order $m$, and $f_{n\ell,0}$ is the corresponding eigenfrequency of the non-rotating star. The quantity $\Omega$ denotes the stellar angular rotation rate, the coefficient $C_{n\ell}$ is the Ledoux constant, and the terms $D_{n\ell}$ and $E_{n\ell}$ represent second-order corrections due to centrifugal distortion and higher-order Coriolis effects, respectively. 
The first-order term $m(1 - C_{n\ell})\Omega$ arises from the Coriolis force and leads to symmetric frequency splitting in a non-magnetic star. The second-order contribution, proportional to $\Omega^2$, includes both a spherically symmetric component $D_{n\ell}$ and an asymmetric component $m^2 E_{n\ell}$, which account for centrifugal distortion and higher-order rotational effects. These second-order terms become significant for moderate to fast rotators and lead to departures from equally spaced multiplets, resulting in asymmetric mode splitting \citep[e.g.][]{Ballotetal2010, Bugnetetal2021, Guoetal2024}. For non-magnetic slow and rigidly rotating stars, Eqn.~(\ref{eqn:ledoux}) reduces to the first-order perturbative regime, defined as: 

\begin{equation}
f_{n\ell m} = f_{n\ell,0} + m \Omega (1 - C_{n\ell}) ~ ,
\label{Eq:first_order}
\end{equation}
such that the symmetric frequency spacing between modes of the same $n$ and $\ell$, but different $m$, becomes approximately:

\begin{equation}
\Delta f \approx m \Omega (1 - C_{n\ell}) ~ .
\end{equation}

 We plotted schematic Fourier spectra of the extracted pulsation frequencies, in which each detected frequency is represented by a vertical line whose height is proportional to its measured amplitude, and first visually examined the spectra for possible rotational splitting. 
 We further performed a semi-automated pair-wise separation computation and a triplet search to identify the rotational split modes. For a multiplet $f_{ijk}$, where $f_{i}$ is the frequency of the retrograde component, $f_{j}$ the frequency of the zonal mode and $f_{k}$ the frequency of the prograde component, an asymmetry parameter, $A_{s} = f_{i} +f_{k} -2f_{j}$ is estimated. The normalised asymmetry parameter is defined as
 
 \begin{equation}
     A_{\mathrm{norm}} = \frac{(f_j-f_i)-(f_k-f_j)}{(f_j-f_i)+(f_k-f_j)}
\label{eqn:assymetry}
\end{equation}

\noindent which was estimated in accordance with \citet{DziembowskiandJerzykiewicz2003} and \citet{Guoetal2024}.
Following \citet{Vanrespailleetal2026}, we adopted a mean splitting of $\Delta f<0.75~{\rm d^{-1}}$  as one of the criteria for identifying rotational multiplets in $\beta$ Cep pulsators. We emphasise that $0.75~{\rm d^{-1}}$ is an empirically motivated selection criterion rather than a sharp physical upper limit on rotational splitting. Frequency multiplets with $0.75<\langle\Delta f\rangle<1.0~{\rm d^{-1}}$ were retained only as exploratory candidates and require additional independent evidence prior to confirmation as rotational multiplets. We therefore use $\langle\Delta f\rangle\leq0.75~{\rm d^{-1}}$ as the conservative splitting criterion for the ensemble analysis and $0.75<\langle\Delta f\rangle<1.0~{\rm d^{-1}}$ as the exploratory criterion.  An upper threshold of $1~\rm d^{-1}$ on the mean splitting was adopted in the exploratory regime because it comfortably exceeds the expected rotational splittings of the targets while remaining sufficiently restrictive to exclude widely separated unrelated frequencies. Candidates in the exploratory regime  are distinguished from the conservative sample throughout the analysis and are not used to determine population-level conclusions.  

In addition to the splitting criterion, we require the frequencies within a multiplet to satisfy $A_{s}\leq 0.1~{\rm d}^{-1}$ and $A_{\rm norm}\leq 0.05$.
While the mean-splitting criterion limits the overall frequency separation,
these asymmetry criteria require the prograde and retrograde components of a multiplet to be approximately
symmetric about the central frequency, reducing the number of configurations
that satisfy the splitting criterion by chance.

\begin{figure*}[htbp]
\centering
\includegraphics[width=0.49\linewidth]{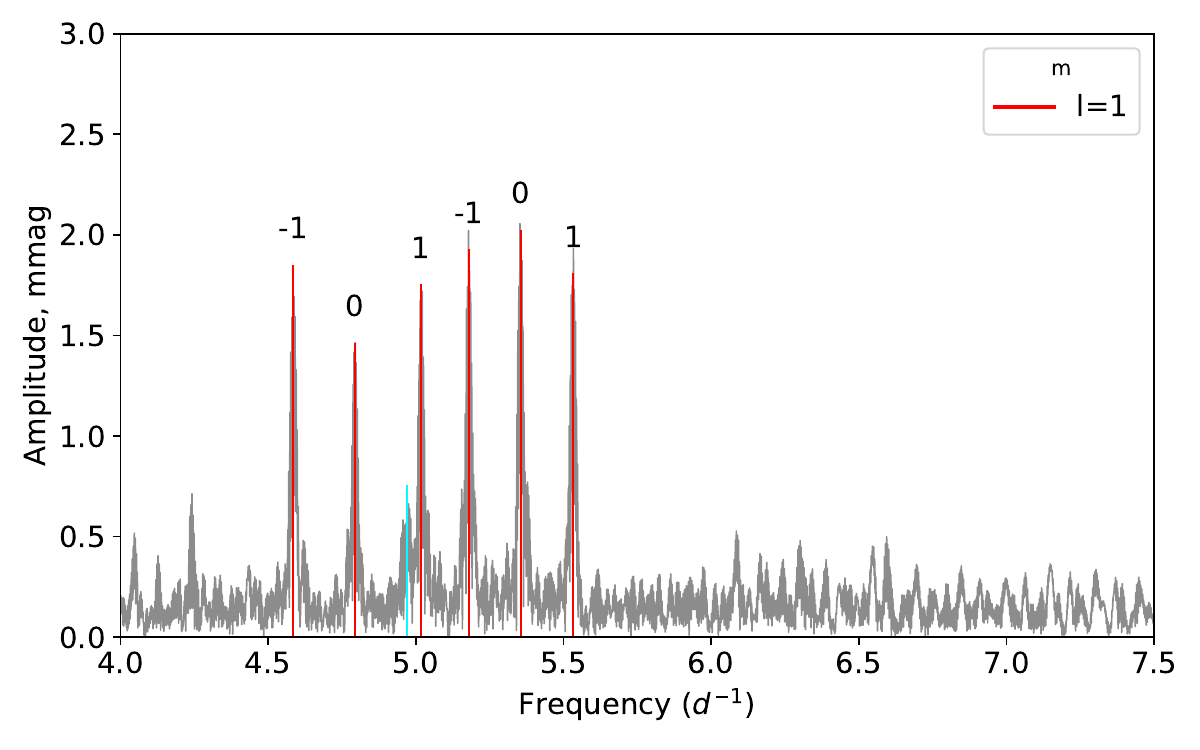}
\includegraphics[width=0.49\linewidth]{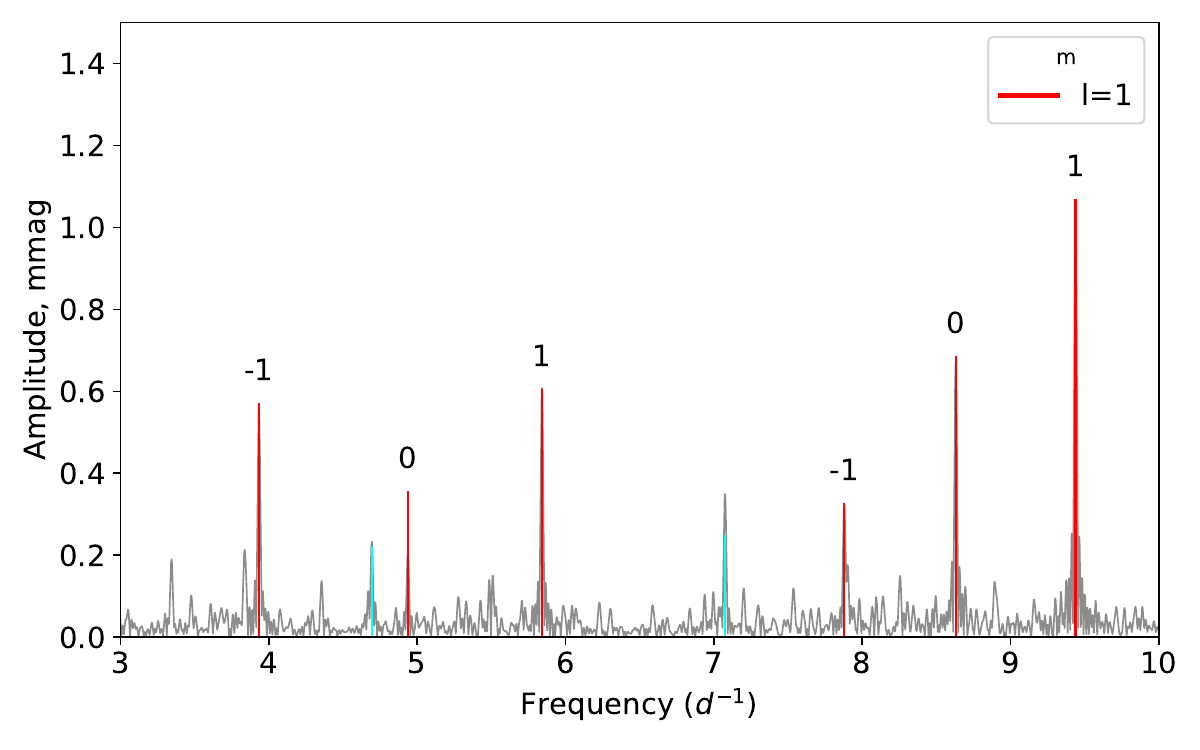}
\caption{Example plots showing the rotational multiplets in TIC355656323 (left panel) and TIC437617380 (right panel).  The schematic spectra are over-plotted onto the continuous Fourier spectra. } \label{fig:fig_rot_split}

\end{figure*}

Table \ref{appendix:tab-rot-multiplet} shows the modes identified in our systems via rotational splitting, and Fig.~\ref{fig:fig_rot_split} shows an example of rotational multiplets for one system. There are also systems that show blended modes, which is a situation where two or more observed pulsation frequencies are so close or overlap, often caused by rotation or low frequency resolution of the data making it difficult to resolve individual components \citep[see e.g.][]{Handler2008, Mirouh2022} . For such systems, we report modes of higher degree $\ell$, or modes of least asymmetry if they are of the same degree. The schematic Fourier plots for rotational multiplets are shown in Figure A.1 in the Appendix. The asymmetries are also shown in Appendix Table \ref{appendix:tab-rot-multiplet}.

Rotational splitting has previously been identified in $\beta$ Cep stars \citep[e.g.][]{DziembowskiandJerzykiewicz2003, Aertsetal2003, Handleretal2004, Briquetetal2007}, which allowed direct mode identification without requiring multicolour photometric data. The splitting also allows direct determination of the interior rotation of stellar systems. Here, splittings are reported in 14 systems with some showing possible mode blends. This blending effect caused by rotation, where, for example, the retrograde component $(-m)$ of a quadrupole mode $\ell =2$ is too close to be resolved from the prograde component $(+m)$ of a set of dipole modes $(\ell =1)$, which could be coincidental or points to a possible geometric frequency shift \citep[e.g.][]{Soufi1998, Daszynskaetal2002}.

The rotational frequencies are estimated from the multiplets  using Eqn.~(\ref{Eq:first_order}). To do that, we estimate the total spacing of a triplet or quintuplet, $\Delta f_{t}$ and divide it by $2m(1 - C_{n\ell})$.
Using the information available from a pre-computed grid of massive stars by Vanrespaille et al. (A\&A, submitted), the term $1 - C_{n\ell}$ approximates to 1 for $\beta$ Cep low-radial order p modes. As a result, we assumed $C_{n\ell}=0$ for the calculation of our rotational frequency. However, since $C_{n\ell}=0$ does not always hold, especially for g and mixed modes, the rotation frequency obtained under such an assumption that involves mixed modes is likely underestimated. Where more than one triplet is observed, we computed the rotational frequency for each triplet and if consistent with a solid rotation, the mean rotational frequency is reported for the system. Otherwise, differential rotation is reported.  The estimated rotational frequencies are shown in Table \ref{appendix:tab-rot-multiplet}.

In the literature, $\beta$~Cep stars are reported to have symmetric and asymmetric mode splittings \citep[e.g.][]{Handleretal2004}. Here, the systems within the adopted asymmetric tolerance level are treated as symmetric splitting. Systems that have asymmetric tolerance beyond the adopted threshold but show clear signatures of splitting are treated as asymmetric. Asymmetric rotationally split mode multiplets have been discussed in detail in \citet{Guoetal2024}. All the multiplets reported here as shown in Table \ref{appendix:tab-rot-multiplet} are treated as symmetric splitting. The dipole modes ($\ell=1$) dominate the identified pulsation modes in the sample. Moreover, the majority of the dominant pulsations in most systems are also identified as dipole modes. However, unidentified peaks exist in the sample, and their number is significantly higher than that of identified ones. In addition, we do not rule out the possibility that some identified triplets are incomplete multiplets of a higher degree $\ell$. Due to the limited number of identified modes in some of the systems, additional mode identification techniques, such as ground-based multicolour photometry
\citep[e.g.][]{Heynderickxetal1994}
or spectral line profile variability
\citep[e.g. see][for a review]{AertsandDeCat2003}

Table \ref{appendix:tab-rot-multiplet} and Fig.~\ref{Appendix:Figa} contain four targets that have two dipole ($l=1$) rotationally split multiplets identified, but the size of their splittings is different (see Figure \ref{Appendix:Figa}). There are several possibilities to understand this. For TIC 437617380 (HD 112485) the two spacings are some 20\% different and the two modes are very different in frequency, hence they likely sample different parts of the stellar interior. Therefore, TIC 437617380 could be rotating differentially. 
The higher frequency mode is more likely to be a pure p mode and shows smaller splitting. Thus, an alternative explanation in terms of mixed modes is rather unlikely. A detailed analysis of this target is presented in Sect.~\ref{sec:asteroseismic_modelling}.  TIC 355656323 (CZ Vel) has splittings with an almost similar structure as TIC 437617380 except that the core and the envelope rotate much slower than those of TIC 437617380.  In TIC 379012185 (EK Cru), the two splittings are different by about a factor of two and the multiplet centroid frequencies are similar, suggesting that if both modes were $l=1$ they would sample similar interior regions of the star and have a similar amount of mode mixing. A steep interior rotation gradient explaining the factor of 2 difference also appears implausible. An alternative explanation is that possibly the triplet with the larger spacing is part of an $l=2$ quintuplet where only the $m \in \{-2, 0, 2\}$ components are observed. This alternative explanation is also supported by the comparison of the asteroseismic model frequencies  with the observed ones, which is discussed in Section \ref{sec:asteroseismic_modelling} and shown in Fig.~\ref{fig:freq_comparison}, where a model frequency of the $\ell =2$ mode coincides with the centroid frequency of the multiplet with the larger spacing. The third case with two apparent dipole multiplets split by rotation is TIC 458599043. However, here the two multiplets with similar centroid frequencies have splittings different by about a factor of 8. For all the arguments mentioned above, we suspect that at least one of these splittings is just a chance agreement.  

In addition, while the zonal mode in TIC 458263480 appears to be clear, it is not certain which frequencies correspond to its non-zonal components. Although the higher amplitude frequencies to the left of its selected $m=\pm1$ components in Fig.~\ref{Appendix:Figa} are cut off  by our tolerance threshold on asymmetry, they also form a plausible triplet with the identified zonal mode with a rotational frequency comparable to that inferred from the selected multiplet.

\section{Rotational synchronisation}\label{sec:synchronisation}
We further investigated the synchronicity or otherwise of the rotation of the stars by comparing their rotation periods inferred from their rotational multiplets with their orbital periods. This provides a direct test of the extent to which tidal torques have modified the stellar rotation and allows us to assess whether each system is close to synchronisation. Tidal interactions, in close binary systems, are expected to drive the stellar spin toward synchronisation with the orbital motion and toward circularisation of the orbit \citep{Zahn1977, Hut1981}. Although both depend on stellar structure, mass ratio, orbital separation, and evolutionary state \citep[e.g.][]{Britavskiyetal2024, Sunetal2026}, they do not always have the same timescales. 

Here, we report the synchronicity or otherwise of the systems with identified multiplets. TIC 134522577 and TIC 379012185 show sub-synchronous rotation, while the rotations of TIC 458599043, TIC 355656323 and TIC 269228628 are fairly synchronised with the orbital motion. TIC 28957011, on the other hand, shows a 2:1 spin-orbit resonance \citep[e.g.][]{Schmidetal2015} and is likely evolving towards a state of  tidal equilibrium. A 2:1 spin-orbit resonance is the situation in which the star rotates twice in one orbital cycle. The rest of the systems show super-synchronous rotation, implying that tidal forces are yet insufficient to synchronise their rotations (see discussion by \citealt{SouthworthandBowman2025}). 

\begin{figure}[htbp]
\centering
\includegraphics[width=\linewidth]{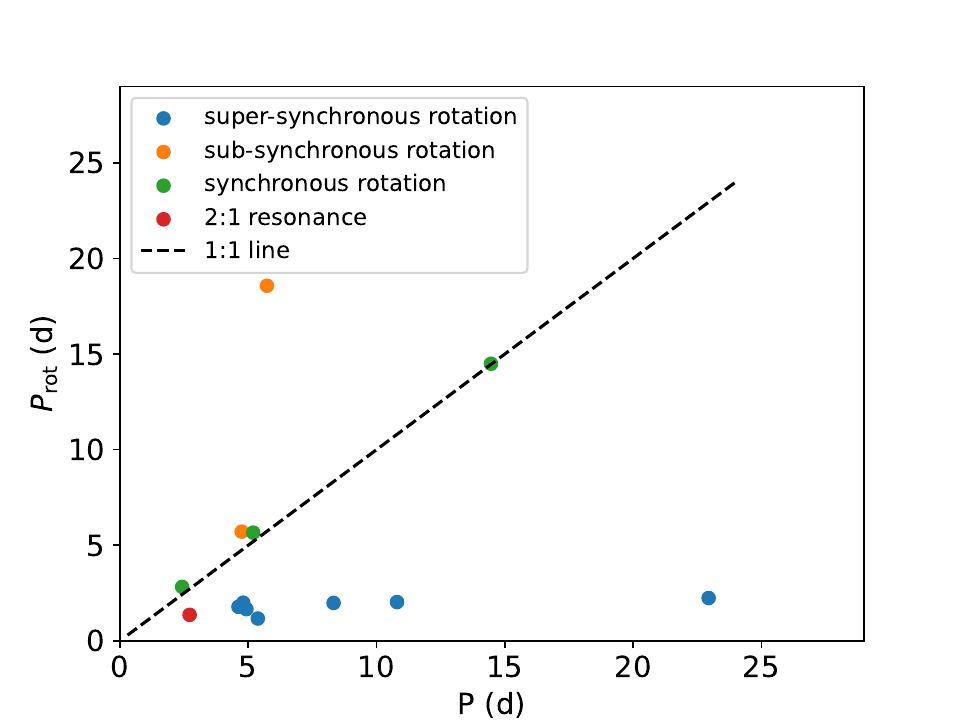}
\caption{Rotation period versus orbital period for systems with detected multiplets except TIC 450918869. TIC 450918869 was excluded from the plot owing to its ambiguous rotation profile. } \label{fig:fig_P-Prot_plot}

\end{figure}

Figure \ref{fig:fig_P-Prot_plot} shows the plot of the rotational period versus the orbital periods.
Our results do not follow strictly  the trends for synchronisation reported in \citet{Lennonetal2024}, for example, in which systems with orbital periods less than $3$~d are expected to be synchronised but those with orbital periods more than $10$~d are expected to be asynchronous. While synchronisation may be more prevalent among close binary systems with short orbital periods indicating the effects of tidal interaction in synchronisation, systems with orbital period greater than $10$~d, TIC 458599043 for example, may also show synchronous rotation. This corroborates the findings that while tidal torque may play a significant role in synchronisation, intrinsic formation processes, efficiency of the angular momentum transport and evolutionary state of the star also contribute to whether a system achieves synchronisation or not \citep[e.g.][]{Levato1976, Tassoul1987, Putkurietal2018, Putkurietal2026}.

To assess the sensitivity of our synchronisation results to the adopted splitting criterion, we repeated the population analysis using maximum spacings of 0.75 and 1.0~d$^{-1}$. The resulting synchronisation fractions  remain approximately 23\% for both criteria in the current sample. Although the more exploratory threshold could admit chance alignments, restricting the analysis to the conservative subsample did not change the synchronisation fraction in the current sample; hence our ensemble conclusion of synchronisation fraction is insensitive to the adopted multiplet-spacing ceiling. Nevertheless, we restricted the  population-level synchronisation statistics to the conservative multiplet sample.

\section{Multivariate Regression Analysis}\label{sec:reg_analysis}

Stellar pulsation amplitudes have been found to be significantly modulated by tidal distortions in binary systems, especially  for tidally trapped or triaxial pulsators \citep{Handleretal2020, Fulleretal2020, Fulleretal2025}. Incidences of  tidally excited modes and tidal frequency shifts have also been reported \citep[][]{FullerandLai2011, Bowmanetal2019d, Preeceetal2019, Guo2021, Fulleretal2025} coupling oscillations and tidal interactions. These modulations and tidally excited oscillations are usually externally induced by centrifugal distortion or corriolis forces. See the review by \citet{SouthworthandBowman2025} for a detailed discussion.

In apparently non-tidally-trapped  pulsators, several relations have also been reported for certain specific  classes of pulsators. Pulsation- and orbital period relationships have been statistically found for $\delta$~Sct stars \citep{Soyduganetal2006, Alicavusetal2017, LiakosandNiarchos2017, Liakos2025}, where orbital periods are found to correlate with pulsation periods. In an ensemble study of $\beta$ Cep stars, \citet{Fritzewskietal2025} reported no significant correlation between pulsation amplitude and frequency, although the amplitude of the dominant pulsation mode appears to show a trend with frequency. The preliminary findings of \citet{Ezeetal2025} showed possible exponential decay trends of the amplitude of dominant pulsations with the pulsation frequency and the orbital period, although the significance was not determined. In addition, we expect that the orbital period in the ensemble would be correlated with the pulsation frequency similar to what is observed for $\delta$~Sct stars \citep{Liakos2025}. However, this behaviour may be different for $\beta$ Cep stars. Again, the possible relations between these parameters may not be monotonic as captured by the univariate regression procedures as there could be possible non-linear interactions between the pulsation and orbital parameters.   

To quantify the relationships between pulsation and binary orbital parameters, we performed Bayesian multivariate regression analyses using the dominant pulsation frequencies. The dominant pulsation frequencies were adopted here as representative pulsations for this sub-analysis, which was also done in earlier studies (e.g. \citealt{Fritzewskietal2025}). 
The identified mode multiplets are predominantly dipole modes. We model the logarithmic pulsation amplitude as a function of pulsation frequency and orbital period using a restricted second-order multivariate regression of the form
\begin{equation}
\log A_i = \beta_0 + \beta_1 \log f_i + \beta_2 \log P_i + \beta_3 (\log f_i)(\log P_i) + \epsilon_i ~ ,
\label{Eq:Reg_eqn}
\end{equation}
where $A_{i}$ and $f_{i}$ are the pulsation amplitude and frequency, respectively, $P_{i}$ is the orbital period, $\beta_0$ is the intercept, $\beta_1, \dots, \beta_n$ are the coefficients and $\epsilon$ is Gaussian noise. The model includes an interaction term between the pulsation frequency and the orbital period to capture potential coupling between these parameters. Due to the limited sample size, we performed the analysis with a minimal number of predictors shown in Eqn.~(\ref{Eq:Reg_eqn}). Because the amplitude distribution is approximately  log-normal, we assumed a normal Gaussian distribution with a Gaussian likelihood and a  weakly informative prior.  If the likelihood is assumed to be 
$y_i \sim \mathcal{N}(\mu_i, \sigma_i^2)$,
where $y_i = \log A_i$,  the mean is given by a linear model
$\mu_i = X_i \beta$,
and the total variance combines intrinsic scatter and observational uncertainty such that
$\sigma_i^2 = \sigma_{\mathrm{int}}^2 + \sigma_{\mathrm{obs},i}^2$. We define the likelihood function explicitly as:

\begin{equation}
\mathcal{L}(y \mid \beta, \sigma_{\mathrm{int}}, X)
=
\prod_{i=1}^{N}
\frac{1}{\sqrt{2\pi \sigma_i^2}}
\exp\left(
-\frac{(y_i - X_i \beta)^2}{2\sigma_i^2}
\right) ~ ,
\label{Eq:likelihood_function}
\end{equation}
\noindent where $\sigma_{\mathrm{int}}$ represents all astrophysical variability that could not be accounted for by the predictors included in the model. 

Adopting a weakly informative prior on the regression coefficients,
$\beta_j \sim \mathcal{N}(0, \tau^2)$, 
and a log-uniform prior on the intrinsic scatter,
$\log \sigma_{\mathrm{int}} \sim \mathcal{U}(-10, 1)$, we defined the posterior distribution as 
\begin{equation}
p(\beta, \sigma_{\mathrm{int}} \mid y, X)
\propto
\mathcal{L}(y \mid \beta, \sigma_{\mathrm{int}}, X)\,
p(\beta, \sigma_{\mathrm{int}}) ~ ,
\label{Eq:probability_distribution}
\end{equation}
where $X$ is the design matrix containing $\log f$, $\log P$, and their interaction term.

Due to the differences in magnitudes of the input parameters, where, for instance, we have high values for frequency and orbital periods but low values for pulsation amplitude, we standardised the input data to ensure quick convergence. The standardisation is such that the data are centred on zero and scaled to unit variance. This improves efficiency, mixing and numerical stability during MCMC run. Posterior sampling was performed using the No-U-Turn Sampler (NUTS) as implemented in PyMC \citep{pymc2023}. 

\begin{table*}
\caption{Posterior summaries of the Bayesian regression parameters, which include the posterior mean, standard deviation, 
94\% highest-density interval (HDI), and Gelman-Rubin convergence statistic ($\hat{R}$).}\label{tab:Multiregmodel}
\centering
\begin{tabular}{lrrrrrrrrrr}

\hline
 & & Mean & SD & $94\%\, \text{HDI}$  &  $\hat{R}$ \\
\hline
Model with interaction & &&&&&&\\
\hline
&Intercept($\beta_0$) & 0.195 & 0.067 & [0.068, 0.323] &  1.0 \\
&Pulsation frequency coefficient ($\beta_1$) & $-0.228$ & 0.139 &[$-0.489$, 0.027] &  1.0 \\
&Orbital-period coefficient ($\beta_2$) & $-0.139$ & 0.379 & [$-0.873$, 0.553] &   1.0 \\
&Interaction coefficient ($\beta_{3}$) & 0.126 & 0.396 & [$-0.590$, 0.904] &   1.0 \\
&Intrinsic scatter ($\sigma_{int}$) & 0.537 & 0.049 & [0.451, 0.631] &   1.0 \\
\hline
Simplified model &&&&&&&\\
\hline
&Intercept($\beta_0$) & 0.195 & 0.067 & [0.068, 0.320] &   1.0 \\
&Pulsation frequency coefficient ($\beta_1$) & $-0.189$ & 0.066 & [$-0.314$, -0.067] &  1.0 \\
&Orbital-period coefficient ($\beta_2$) & $-0.021$ & 0.067 & [$-0.146$, 0.105] &   1.0 \\
&Intrinsic scatter ($\sigma_{int}$)  & 0.533 & 0.049 & [0.445, 0.626] &   1.0 \\ 
\hline
\end{tabular}
\end{table*}

Table \ref{tab:Multiregmodel} shows the resulting statistical solutions with a posterior predictive $p$-value of $\approx$ 0.51 for the models (see \citealt{Gelmanetal1996}). From Table \ref{tab:Multiregmodel}, it can be seen that the posterior means of the regression coefficients associated with pulsation frequency, orbital period, and their interaction are poorly constrained with very large uncertainties and wide HDI (Highest Density Interval). In contrast, the intercept and the intrinsic scatter are well determined, with the latter suggesting that additional physical factors beyond those included in the model contribute to the observed variability in pulsation amplitudes. No significant correlation is observed between the  pulsation amplitude or frequency  and the orbital period. This is not unexpected since the periodicities of the low-radial order p modes in $\beta$ Cep stars are not sub(multiples) of the orbital period. Hence, achieving mode-tidal resonances is harder in this class of pulsations than for higher order g modes where the orbital period and mode periodicities are in the same region of timescales, namely, both of order of days \citep[see][]{AertsandTkachenko2024}. However, an indirect dependence could arise through binary evolution. Mass transfer may rejuvenate the mass-accreting component in close binary systems, modifying its internal structure and potentially shifting its pulsation frequencies (see \citealt{SouthworthandBowman2025}). If such effects are relevant in the current sample, one might expect systematically different pulsation properties among systems with shorter orbital periods. The absence of a significant trend in the present sample may also suggest that such evolutionary effects are weak, not present in most systems, or are masked by the diversity of stellar masses, evolutionary states, and binary interaction histories represented in the sample.

\begin{figure*}[htbp]
\centering
\includegraphics[width=0.45\linewidth]{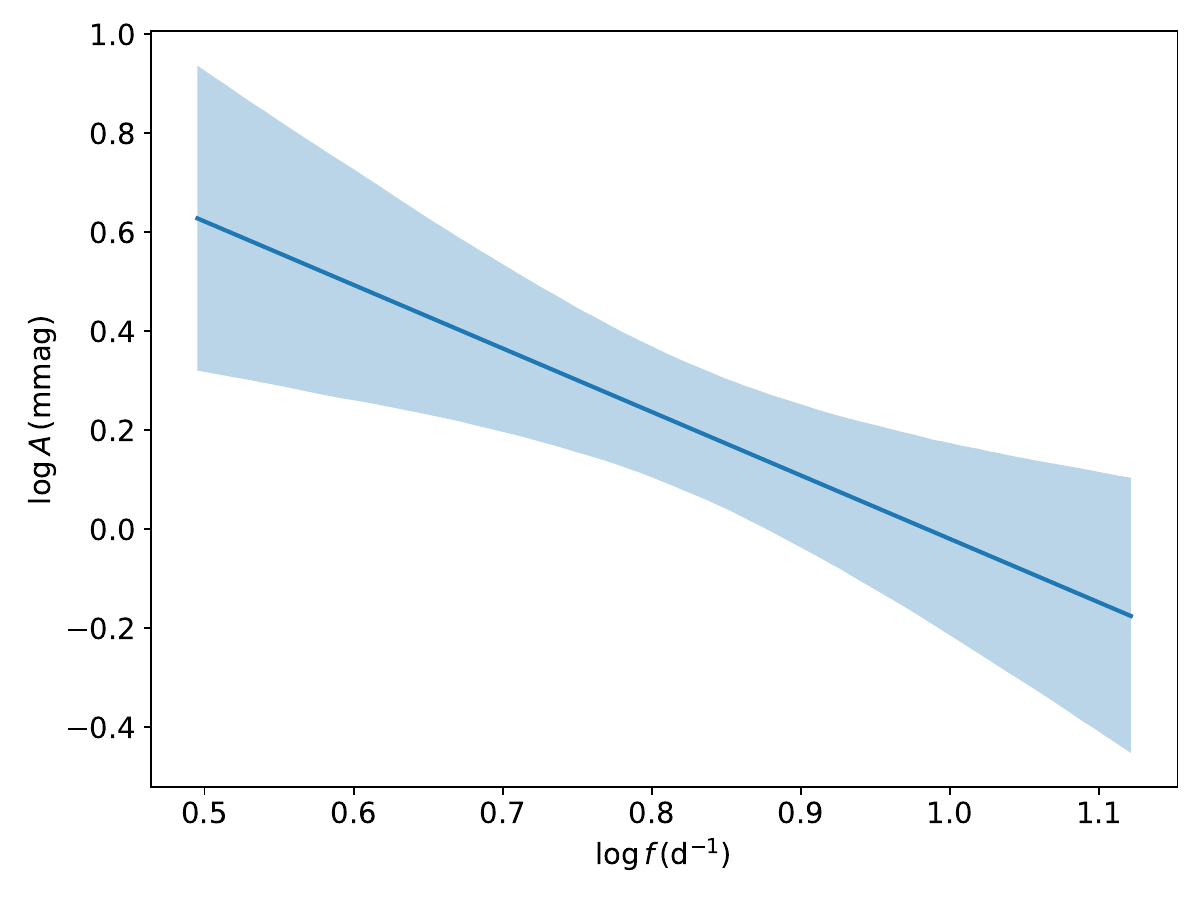}
\includegraphics[width=0.45\linewidth]{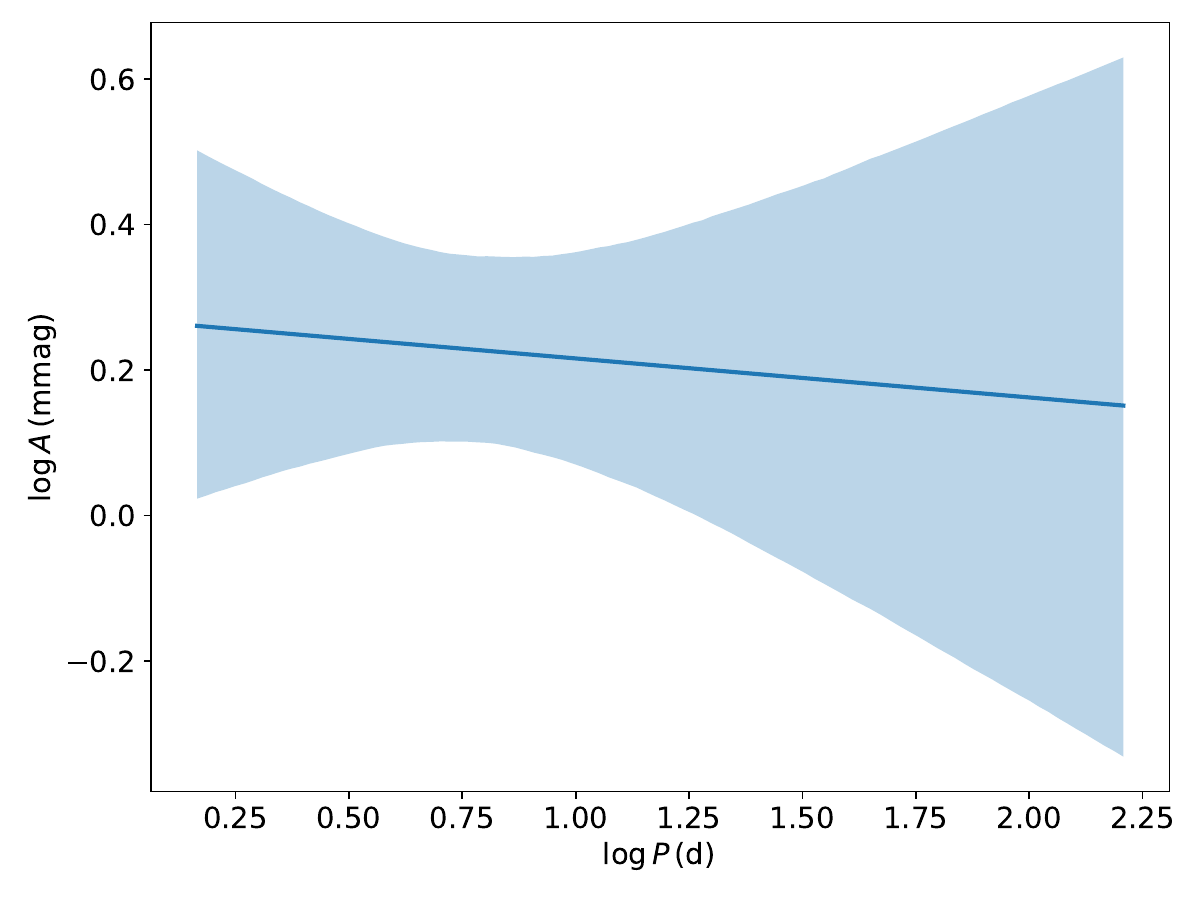}
\caption{Conditional relationship between pulsation amplitude and frequency at fixed orbital period (left panel) and orbital period at fixed pulsational frequency (right panel). The solid lines represent the posterior mean relation, while the shaded regions denote the corresponding 94\% credible interval.}\label{fig:simplified_model}

\end{figure*}

Comparing the complete model to the simplified model obtained under similar conditions but without the interaction term ($\beta_3 (\log f_i)(\log P_i)$), using the leave-one-out-cross-validation (LOO), the simplified model is preferred and has a well constrained negative amplitude dependence on the pulsation frequency, but not with the orbital period. The statistical evidence that higher-frequency modes exhibit lower amplitudes is consistent with the observed statistical distribution. The simplified regression model is therefore adopted as the primary model, while the complete model with interaction term is retained as a sensitivity analysis. To visualise the individual dependencies of pulsation amplitude on pulsation frequency and orbital period in the simplified model, we constructed conditional-effect plots from the posterior distributions. These are shown in Fig.~\ref{fig:simplified_model}. For each predictor, the posterior expected value of $\log A$ was evaluated across the observed range of that predictor while holding the other predictor fixed at its sample median. These plots illustrate the model-inferred dependence of the expected pulsation amplitude on each predictor after accounting for the contribution of the other predictor. Additionally, the posterior mean 2D response surface map and uncertainty shown in Fig.~\ref{fig:fig_model_with_interaction} for the model with the interaction term, as well as the univariate regression analysis shown in Fig.~ \ref{fig:appendix_univariate_reg} obtained under similar conditions with pair-wise use of the parameters, corroborate the findings visualised in the simplified model.

The findings in this paper on the relation between pulsation and orbital period for $\beta$ Cep stars do not agree with reports in the literature for $\delta$~Sct stars \citep[e.g.][]{Soyduganetal2006, Alicavusetal2017, LiakosandNiarchos2017, Liakos2025} and the modulation of the amplitude by orbital interaction is also not  statistically significant within our sample. Although the models in this analysis capture some trends in the data, the increasing uncertainty at long orbital periods and the poor constraints on the parameters in the model with interaction terms highlight the need for larger samples and improved observational constraints. Future work could incorporate additional physical parameters, such as stellar mass, rotation rate, and metallicity, which are still very limited for $\beta$ Cep stars in eclipsing binaries.  Such extensions would enable a more comprehensive understanding of the interplay between pulsation, rotation, and binarity in massive stars and would investigate the parameters that contribute to intrinsic variability further.

\section{Selected cases of asteroseismic modelling via rotational splitting}\label{sec:asteroseismic_modelling}
We conducted the asteroseismic analysis of two of the candidates (HD 112485 and EK Cru) that show rotational multiplets using \textsc{Ampharos} (Vanrespaille et al. (submitted)) and its accompanying model grid. \textsc{Ampharos} uses \text{MESA} \citep{Paxtonetal2011, Paxtonetal2013, Paxtonetal2015, Paxtonetal2018, Paxtonetal2019, Jermynetal2023} evolutionary and \textsc{StORM} (Vanlaer et al. submitted) pulsation grids of models of varying initial masses, convective boundary mixing with a diffusive exponential prescription, envelope mixing, core hydrogen mass fraction, and rotation to perform the asteroseismic analysis. The grid was computed for a range of $7-29.85\, M_{\odot}$ for mass, $1-6 \, \mathrm{cm^{2}~s^{-1}}$ for $\log D_\mathrm{mix}$, $0.005-0.035$ for $f_\mathrm{ov}$, $0.0001 -0.701$ for $X_{c}$. and $0-0.40$ for $\frac{f_{\rm rot}}{f_{\rm crit}}$. While the \textsc{MESA} models are non-rotating, the \textsc{StORM} grid captures the rotation. The latter also captures the distortions to the geometry of the star arising from rotation using the Chandrasekhar-Milne approximation to second order in rotation frequency \citep{Chandrasekhar1933, Tassoul1978}. Similar to the approach taken by \citet{Burssensetal2023}, \textsc{Ampharos} minimises a $\chi^{2}$ merit function of a subset of models with a fixed age and rotation frequency that optimally reproduce a favoured zonal mode and the rotational splitting in a star. 
Its merit function considers the luminosity, effective temperature, identified zonal mode frequencies, and identified rotational splittings. From a weighted average based on the merit function over the subset of models the asteroseismic stellar parameters and interior properties of the stars are derived. 

Our selected targets are reported as detached eclipsing double-lined spectroscopic (SB2) binaries (\citealt{Cakirlietal2025}, Eze et al. (A\&A, submitted)). Their detached nature allowed us to use the grid of models built for single stars. The selected targets have available stellar parameters and at least two identified modes. Their luminosity and effective temperature used for the analysis were obtained from Eze et al.\ (submitted) and \citet{Cakirlietal2025}. The identified multiplets in both systems are rotationally split triplets of $\ell= 1\, \text{and}\, 2$ modes, leading to the use of a non-radial 
mode setting in \textsc{Ampharos}. 
We first searched for the best fit models in the entire $\beta$ Cep instability strip that reproduce the observed frequency using a radial order range of $n_{pg} \in \{-3, -2, ..., +3\}$. In each of the models, a unique $n_{pg}$ that is consistent with the observed zonal modes of the multiplets and the spectroscopic constraints is identified. The identified models are then refitted within a subset of the model grid compatible with the observed spectroscopic parameters.

The uncertainties were estimated from the ensemble of acceptable stellar models obtained during the forward seismic modelling. Models satisfying the adopted spectroscopic constraints were retained and their parameter distributions examined. The likelihood weighting based on $\chi^2$ values obtained from the forward asteroseismic modelling resulted in an effective sample size of approximately one model, producing unrealistically small formal uncertainties. As a result, the final confidence intervals were derived from the unweighted distributions of the accepted models. For each parameter, the lower and upper uncertainties correspond to the 16th and 84th percentiles of the accepted-model ensemble, respectively. Thus, these intervals represent the range of stellar properties consistent with the seismic and spectroscopic constraints within the explored model grid.   

\begin{figure*}[htbp]
\centering
\includegraphics[width=0.49\linewidth]{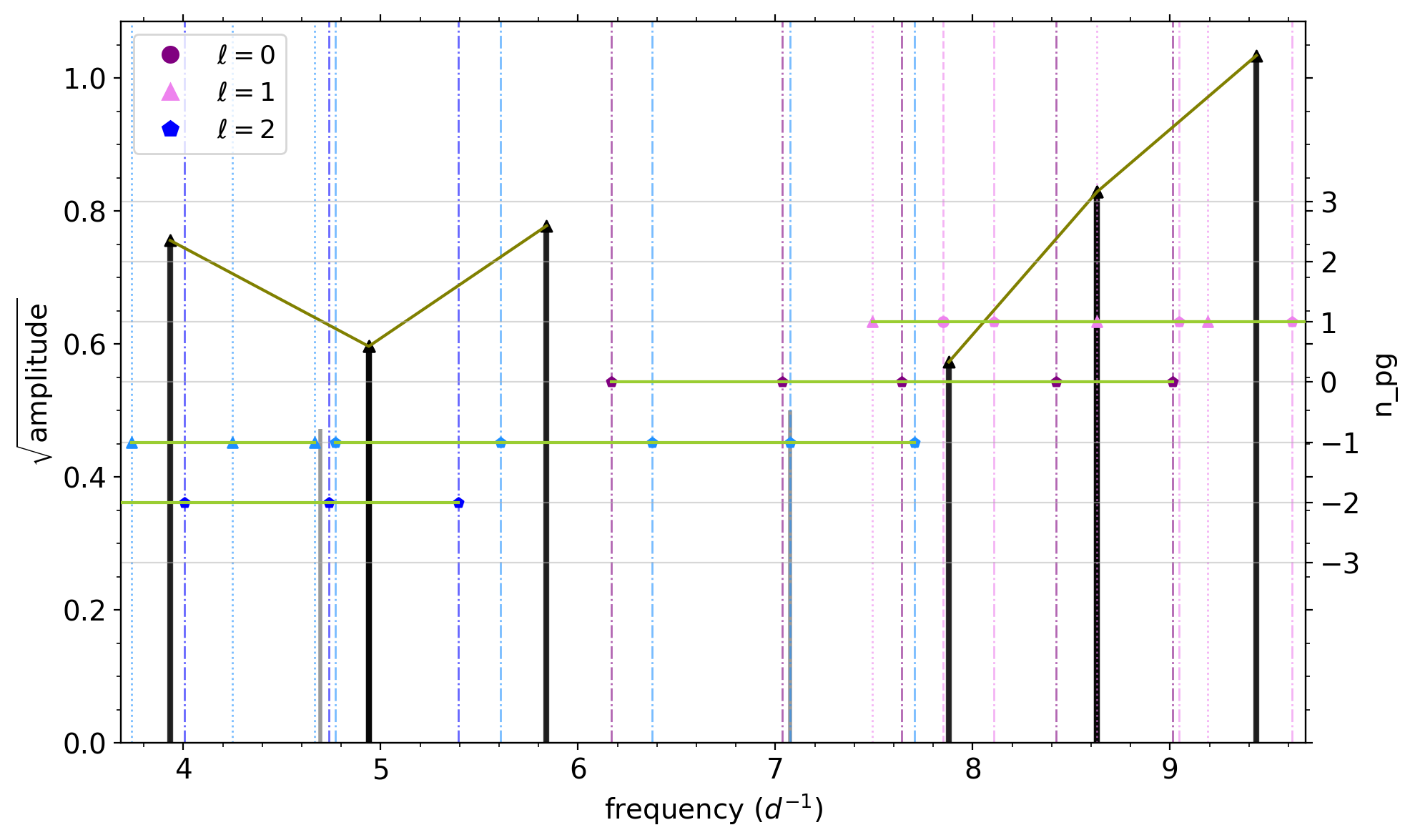}
\includegraphics[width=0.49\linewidth]{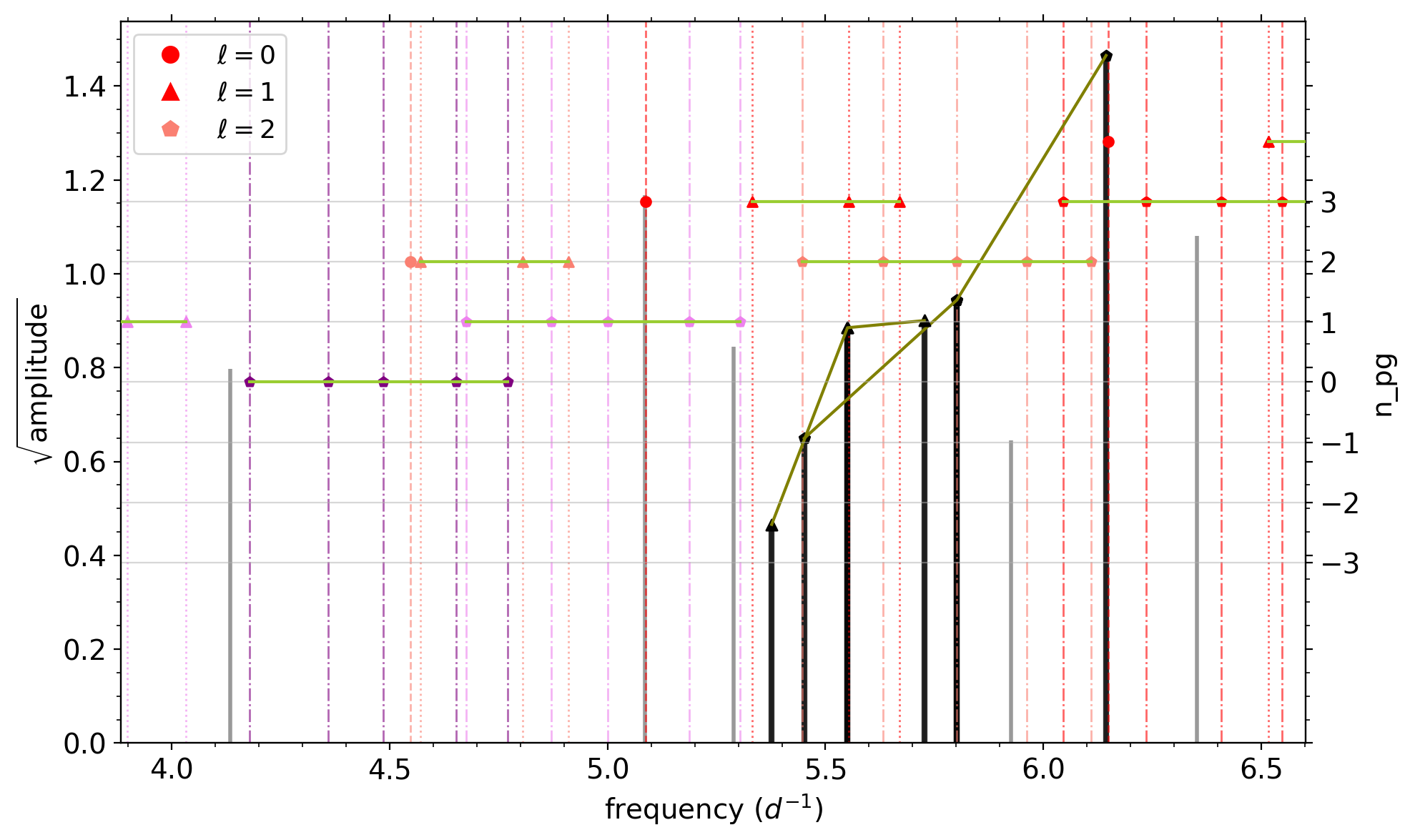}
\caption{Comparison of observed and best-fit model pulsation frequencies of HD 112485 (left panel) and EK Cru (right panel). Identified modes are in black and the unidentified modes are in gray. Other coloured lines denote model mode frequencies of different angular degrees. Triangles denote dipole modes, diamonds denote quadrupole modes and circles denote radial modes. }\label{fig:freq_comparison}

\end{figure*}

The derived asteroseismic parameters of the targets and their uncertainties are shown in Table \ref{tab:asteroseismic_param} and the comparison of the observed and seismic model frequencies are shown in Fig.~\ref{fig:freq_comparison}. Figure \ref{fig:appendix_unweighted_distribution} shows an example plot of the unweighted distributions of mass and radius. As seen in Fig.~\ref{fig:freq_comparison}, the observed zonal modes are well reproduced by the models. The two reproduced zonal modes in HD 112485 are the first-order dipole p modes $-$ sensitive to the envelope $-$ and the second-order quadrupole g mode $-$ sensitive to the near-core structure $-$, respectively, enabling the characterisation of the entire star. 
The modes of EK Cru are 
dipole and quadrupole modes. The derived asteroseismic stellar parameters agree fairly with the observed parameters of the targets \citep{Cakirlietal2025, Nazeetal2025}, Eze et al. (submitted). \citet{Nazeetal2025} and Eze et al. (submitted) obtained a mass of $11.6\pm0.6\,M_{\odot}$ and $11.3_{1.3}^{1.4}~\rm M_{\odot}$, respectively, for HD 112485, while \citet{Cakirlietal2025} obtained a mass of $22.68\pm0.04~\rm M_{\odot}$ for EK Cru.

We obtain convective core masses of $3.30_{-0.21}^{+0.19}~\rm M_{\odot}$  and $8.2_{-0.6}^{+0.5}~\rm M_{\odot}$ for HD 112485 and EK Cru, respectively, resulting in core-mass to total mass fractions of 0.281 and 0.373, respectively. These agree with the possible size of the convective core masses for systems of similar masses in the literature \citep[e.g.][]{Tkachenkoetal2020}. The relative mass of the convective core is affected by mass loss, interior mixing, and convective boundary mixing among other processes, and these physical processes increase the relative mass fraction of the core \citep[e.g.][]{Maeder1986, Kaiseretal2020, Shikauchietal2025, Whiteheadetal2026}. Since the core mass is larger for larger stellar mass and/or larger CBM, the higher relative core mass obtained for EK Cru compared to HD 112485 is expected.

The core hydrogen mass fractions are $0.501_{-0.013}^{+0.009}$ and $0.271_{-0.037}^{+0.034}$, respectively, for HD 112485 and EK Cru, with HD 112485 being less evolved than EK Cru. Although both are of a comparable age, EK Cru has evolved about twice as much compared to HD 112485, which is about half as massive and  consistent with the established fact that more massive stars evolve faster. The inferred central hydrogen mass fraction, $X_{c} = 0.501_{-0.013}^{+0.009}$ reported here for HD 112485 is larger than that reported for any  $\beta$ Cep pulsators subjected to detailed seismic modelling \citep[e.g.][]{Pamyatnykhetal2004, Dupretetal2004, Briquetetal2007,Handleretal2009, Aertsetal2011, Briquetetal2012, Daszynskaetal2013, Burssensetal2023, Vanlaer2025}, suggesting that the star is the least evolved $\beta$ Cep stars modelled to date.

The rotation profiles of the models in the grid used by $\textsc{Ampharos}$ are assumed to be uniform (solid-body). Therefore, the rotational frequencies reported for the targets in Table \ref{tab:asteroseismic_param} effectively correspond to the best-fitting rotation rates of the models. 

\renewcommand{\arraystretch}{1.5}
\begin{table}[htbp]
\begin{center}
\caption{Derived asteroseismic model parameters of two $\beta$~Cep stars.}\label{tab:asteroseismic_param}
\begin{tabular}[t]{lcccl}
\hline
Star &	HD 112485 &	EK Cru\\
\hline
$M\, (\rm M_{\odot})$	& $11.75_{-0.41}^{+0.43}$ & $21.88_{-0.75}^{+1.02}$\\
$M_{\rm cc} \, (\rm M_{\odot})$ &	$3.30_{-0.21}^{+0.19}$ &  $8.154_{-0.603}^{+0.504}$\\
$\frac{M_{\rm cc}}{M}$	& $0.28_{-0.01}^{+0.01}$ & $0.373_{-0.029}^{+0.028}$\\
$R\, (\rm R_{\odot})$	& $5.191_{-0.056}^{+0.057}$ & $11.206_{-0.322}^{+0.179}$\\
$X_c$	& $0.501_{-0.013}^{+0.009}$ & $0.271_{-0.037}^{+0.034}$\\
$f_\mathrm{ov}$ &	$0.005_{-0.010}^{+0.010}$ & $0.03_{-0.01}^{+0.02}$\\
$f_{\rm rot}\,({\rm d}^{-1})$ &	$0.882_{-0.008}^{+0.008}$ & $0.175_{-0.003}^{+0.002}$\\
$\frac{f_{rot}}{f_{crit}}$ &	$0.3508_{-0.0004}^{+0.0005}$ & $0.151_{-0.1}^{+0.1}$\\
$\log(D_{\rm mix}/\mathrm{cm^2\,s^{-1}})$ &	$1.0_{-2.0}^{+2.0}$ & $2.35_{-2.00}^{+1.00}$\\
$\log (L/\rm L_{\odot})$ & 	$4.08_{-0.04}^{+0.04}$ & $5.014_{-0.024}^{+0.032}$\\
$\log (T_{\rm eff}/{\rm K}$) &	$4.422_{-0.006}^{+0.007}$ & $4.489_{-0.006}^{+0.009}$\\
$\log (\mathrm{age}/{\rm yr})$ &	$6.89_{-0.05}^{+0.04}$ & $6.819_{-0.032}^{+0.030}$\\
$\log g\, ({\rm dex})$ &	$4.078_{-0.006}^{+0.006}$ & $3.677_{-0.009}^{+0.023}$\\
\hline
\end{tabular}
\end{center}
\end{table}
\renewcommand{\arraystretch}{1.0}

\subsection{Assessing differential rotation in HD 112485}
The candidate multiplets of HD~112485 deserve particular attention because their mean splitting is in our adopted exploratory regime of $0.75 - 1.0~{\rm d^{-1}}$. We therefore do not regard these frequency splittings alone as sufficient evidence for a rotational-multiplet interpretation. In addition to deriving global stellar parameters, the forward asteroseismic model gives a clear picture of how well the model reproduces the observations. As seen in Fig.~\ref{fig:freq_comparison} (left panel), the zonal mode of the higher frequency  multiplet is well reproduced, but the zonal mode of the lower frequency multiplet is less so. Yet, it is consistent with a second-order quadrupole g mode. Additionally, two unidentified observed mode frequencies, 4.6935(8) and 7.0730(8)~d$^{-1}$, are also reproduced as prograde components of a triplet and quintuplet, respectively. Consequently, we adopt the outer multiplet as the more secure identification, while the inner lower-frequency multiplet is retained as a candidate.  

Due to the rigid body rotation implemented in \textsc{AMPHAROS} and \textsc{StORM}, each stellar structure is associated with a single rotation frequency rather than with a radial differential rotation profile. As a result, we cannot directly measure the exact shape of the rotation profile. However, we can test whether the two identified multiplets are consistent with the same effective rotational frequency or not. This allows us to interpret the difference (if any) as evidence for radial differential rotation. To do this, we treated each multiplet as an independent rotational diagnostic while retaining the same underlying stellar structure. For a given mode identification $(n_{\rm pg},\ell,m)$, \textsc{StORM} provides the rotationally modified frequencies $f_{n\ell m}(f_{\rm rot})$ over a discrete rotation grid. For each mode, we retain the portion of the StORM track over which the same mode can be consistently identified and followed as $f_{\rm rot}$ varies, and interpolate its frequency as a function of $f_{\rm rot}$. The comparison is then restricted to the rotation-frequency range common to all components of each multiplet.

At each trial $f_{\rm rot}$, the corresponding model splittings were
calculated directly from the interpolated StORM frequencies and compared with the observed splitting.  The effective rotation frequency for each multiplet was obtained by minimising a $\chi^2$ merit function that includes the observed and modelled splittings, and accounts for their observational uncertainties and the covariance arising from their common $m=0$ component. The uncertainties on $f_{\rm rot}$ were derived from the $\Delta\chi^2=1$ profile for one fitted parameter. The effective rotation frequencies are interpreted as being associated with the regions probed by the individual multiplets in the best-fitting non-rotating forward asteroseismic structure model. A difference between effective rotation frequencies may indicate non-uniform rotation, but does not by itself constrain the radial rotation profile.

The high-frequency multiplet centred at
$f \simeq 8.63\,{\rm d}^{-1}$ is identified as a dipole p mode with an
effective rotational frequency of
$0.857_{-0.003}^{+0.003}$~d$^{-1}$ and is predominantly
sensitive to the envelope. The lower-frequency multiplet may form part of a
quintuplet. However, an $m=\{-2,0,+2\}$ identification yields
$f_{\rm rot}=0.5316_{-0.0001}^{+0.0001}$~d$^{-1}$, which is lower than the
envelope-sensitive estimate and therefore unlikely, as most of the $\beta\,$Cep stars with inferred internal rotation reveal a decreasing rotation rate from deeper internal layers towards the outer envelope \citep{Vanrespailleetal2026}. We consequently favour the interpretation that
the observed components correspond to $m=\{-1,0,+1\}$, for which
$f_{\rm rot}=0.997397_{-0.000034}^{+0.000006}$~d$^{-1}$. This value lies
near the upper edge of the \textsc{StORM} rotation grid and should therefore
be treated with caution. At this rotation rate, the predicted retrograde
and prograde splittings are $0.9276$ and $0.8158$~d$^{-1}$, differing from
the observed values by $0.0786$ and $0.0854$~d$^{-1}$, respectively.
The observed and modelled splitting asymmetries are comparable. 
Moreover, the ratio 0.997/0.857=1.164 is in line with the measured upper limits for the internal versus surface rotation rates deduced by \citep{Fritzewskietal2025} for their sample of $\beta\,$Cep pulsators.

We emphasise that no radial differential rotation profiles are included in the pulsation modelling. The sensitivity kernels are used only to
interpret the radial regions probed by the modes and do not enter the
optimisation of $f_{\rm rot}$. Hence, differences between the inferred
effective rotation frequencies may indicate radial differential rotation,
but do not deliver its radial profile.

\subsection{Rotational sensitivity Kernel}
To check the sensitivity regions of the inferred rotation frequencies inside HD~112485 and EK~Cru, we estimated the rotational kernel using \textsc{StORM} and the derived asteroseismic forward model of each target. 
As common in forward asteroseismic modelling \citep[e.g.,][]{Aerts2021}, the rotational sensitivity kernel was not included as part of the optimisation procedure of
\textsc{AMPHAROS}. Rather, we rely on it in our current post-processing while 
using the optimised stellar structure model for the derived parameters from \textsc{AMPHAROS}. 

\begin{figure*}[ht]
\centering
\begin{subfigure}[t]{.43\textwidth}
\centering
\begin{overpic}[width=\linewidth]{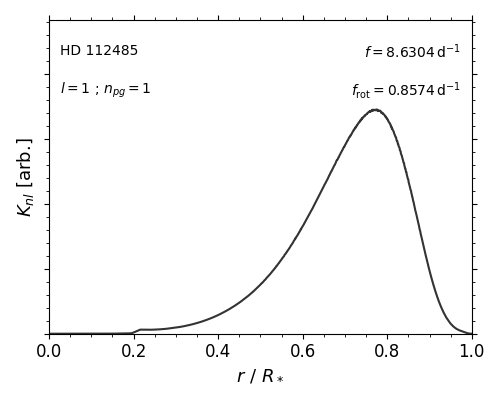}
  \put(20, 50){\small\bfseries }
\end{overpic}
\caption{}\label{fig:cpd45}
\end{subfigure}
\begin{subfigure}[t]{.43\textwidth}
\centering
\begin{overpic}[width=\linewidth]{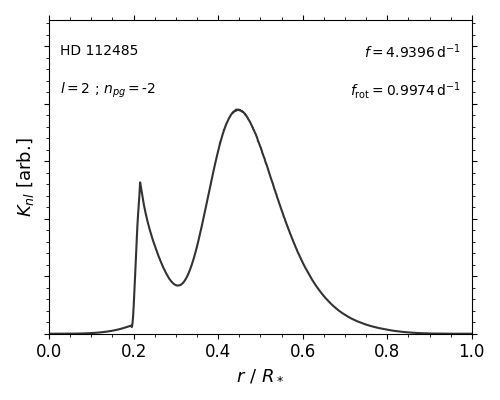}
  \put(10, 50){\small\bfseries }
\end{overpic}
\caption{}\label{fig:hd101}
\end{subfigure}

\medskip

\begin{subfigure}[t]{.43\textwidth}
\centering
\begin{overpic}[width=\linewidth]{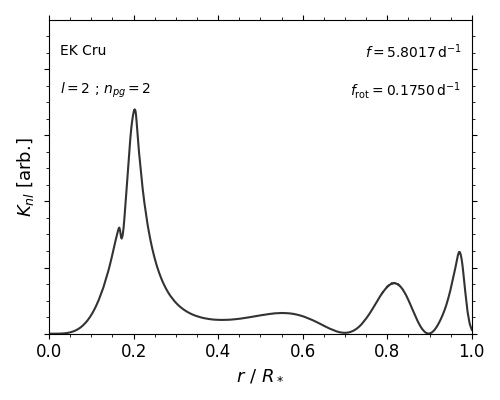}
  \put(10, 50){\small\bfseries }
\end{overpic}
\caption{}\label{fig:hd108}
\end{subfigure}
\begin{subfigure}[t]{.43\textwidth}
\centering
\begin{overpic}[width=\linewidth]{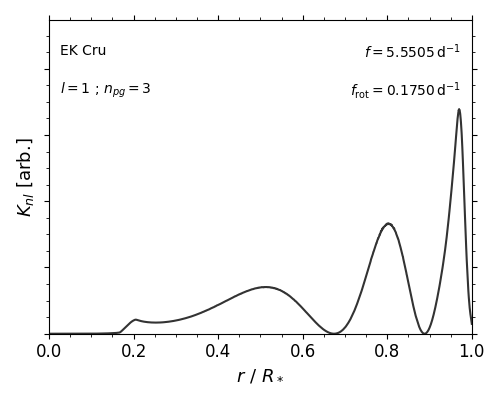}
  \put(30, 50){\small\bfseries }
\end{overpic}
\caption{}\label{fig:hd1124}
\end{subfigure}

\caption{Plots showing the rotational sensitivity kernels of HD 112485 (upper panel) and EK Cru (lower panel), respectively.}\label{fig:rot_sensitivity_kernel}
\end{figure*}

The rotational sensitivity kernel \citep{Aertsetal2010} is defined as:
\begin{equation}
K_{n\ell}(r) =
\frac{
\rho(r)\, r^2 \left[
\xi_r^2(r) + \ell(\ell+1)\,\xi_h^2(r)
- 2\,\xi_r(r)\,\xi_h(r) - \xi_h^2(r)
\right]
}{
\int_0^R \rho(r)\, r^2 \left[
\xi_r^2(r) + \ell(\ell+1)\,\xi_h^2(r)
\right] \, \mathrm{d}r
}\, ,
\label{eq:rot_kernel}
\end{equation}
where $r$ denotes the radial coordinate, $R$ the stellar radius, and 
$\rho(r)$ the density profile. The quantities $\xi_r(r)$ and 
$\xi_h(r)$ are the radial and horizontal components of the displacement 
eigenfunction, respectively. 
\textsc{StORM} is used to compute the radial and horizontal displacements.
The sensitivity kernels for the two considered stars are shown in Fig.~ \ref{fig:rot_sensitivity_kernel} for HD 112485 (upper panel) and EK Cru (lower panel).

The rotational kernels of HD~112485 indicate that the modes associated with the two multiplets have different radial sensitivity distributions. The higher frequency mode multiplet in HD~112485 with the zonal mode frequency $f=8.6304$~d$^{-1}$ 
and rotational frequency of $0.857_{-0.003}^{+0.003}$~d$^{-1}$ 
is sensitive to a broad part of the envelope of the star, which is characteristic of p modes. The lower-frequency mode ($f=4.9396$~d$^{-1}$) has predominantly a g-mode character, with substantial sensitivity to the near-core and innermost half of the model and gives a rotational frequency of $0.997397_{-0.000034}^{+0.000006}$~d$^{-1}$. The difference in their inferred effective rotation frequencies, together with their distinct radial sensitivities, therefore provides tentative evidence of differential rotation. Yet, this interpretation requires future verification from actual dedicated rotation inversions for the star because the zonal mode of the lower-frequency multiplet is less well reproduced and its preferred rotation solution lies near the upper limit of the available \textsc{StORM} grid.  

In EK Cru, the frequency of the zonal mode, $f=5.8017$~d$^{-1}$, is sensitive to both the near-core region and the envelope, characteristic of a mixed mode. It is a low-radial order mixed mode with a substantial g-mode character and appreciable sensitivity to the outer envelope, and thus samples both the core and the envelope, while $f=5.5505$~d$^{-1}$ samples only the envelope.  Both the core and the envelope have similar rotational frequencies and are compatible with a quasi-rigid rotation for the primary of this system. 

\section{Conclusions}\label{sec:conclusion}
We conducted an ensemble pulsation study of a sample of 65 $\beta$\,Cep pulsators in eclipsing binaries, as well as in-depth asteroseismic analysis of two selected cases in the sample. We determined the distribution of the pulsational properties, identified the pulsational mode characteristics, inferred the statistical relations between the pulsation and binary parameters, and derived the asteroseismic parameters of HD~112485 and EK~Cru. 

The distributions of the pulsation frequencies and amplitudes are unimodal and positively skewed. The identified modes are predominantly dipole modes, with some systems showing synchronous, 2:1 resonance, sub-synchronous, and super-synchronous rotation rates. HD~112485 shows tentative evidence for radial differential rotation, while the other systems with identified multiplets, except CZ~Vel, are consistent with quasi-rigid rotation. Although forward asteroseismic modelling was not conducted here for CZ~Vel, the effective rotation frequencies inferred from its two multiplets appear to differ. The inner multiplet indicates a rotation frequency approximately twice the orbital frequency, whereas the outer multiplet indicates a rotation frequency slightly lower than the orbital frequency. A detailed analysis of the internal rotation of CZ~Vel will be presented in our follow-up work. The pulsation amplitude of the systems in the sample appears to have an anti-correlation with the pulsation frequency but shows  no significant correlation with the orbital period. The frequency also shows no significant relation to the orbital period in the current sample. The interior properties of two $\beta$~Cep stars were asteroseismically derived, yielding convective core masses of $3.30\, \rm M_{\odot}$ for the $11.75\, \rm M_{\odot}$ star HD~112485 and $8.154\, \rm M_{\odot}$ for the $21.88\, \rm M_{\odot}$ star EK~Cru. The evolutionary states, interior mixing, and effective rotation frequencies and tentative constraints on radial differential rotation are reported.

\begin{acknowledgements}
This work was supported by the Polish National Science Foundation (NCN) under grant nr. 2021/43/B/ST9/02972 led by GH. 
CIE thanks the Polish National Agency for Academic Exchange (NAWA) for funding, through the Bekker NAWA grant number BPN/BEK/2024/1/00049, the internship where part of this work was conducted. The author also thanks A. Miszuda for helpful comments.
DMB gratefully acknowledges UK Research and Innovation (UKRI) in the form of a Frontier Research grant under the UK government's ERC Horizon Europe funding guarantee (SYMPHONY; PI Bowman; grant number: EP/Y031059/1), and a Royal Society University Research Fellowship (PI Bowman; grant number: URF{\textbackslash}R1{\textbackslash}231631). 
CA and AK acknowledge financial support from the Flemish Government under the long-term structural Methusalem funding programme by means of the project SOUL: Stellar evolution in full glory, grant METH/24/012. 
MV received funding from the KU Leuven Research Council (doctoral mandate grant DB/24/008).
MV and CA 
acknowledge financial support from the European Research Council (ERC) under the Horizon Europe programme (Synergy Grant agreement N◦101071505: 4D-STAR). While partially funded by the European Union, views and opinions expressed are however those of the authors only and do not necessarily reflect those of the European Union or the European Research Council. Neither the European Union nor the granting authority can be held responsible for them. AT acknowledges support from the BELgian federal Science Policy Office (BELSPO) through PRODEX grant PLATO (ZKE8588), and from the Research Foundastion – Flanders (FWO) (grant agreement G0ABL24N).
This paper used the TESS data obtained from the Mikulski Archive for Space Telescopes (MAST) at the Space
Telescope Science Institute. 
Support to MAST for these data is provided by the NASA Office of Space Science via grant NAG5-7584 and by other grants and contracts. Funding for the TESS mission is provided by the NASA Explorer Programme.

\end{acknowledgements}

 \bibliographystyle{aa} 
\bibliography{references} 

\begin{appendix}
\onecolumn

\section{Rotational multiplets of $\beta$ Cep eclipsing binaries  in TESS data}

Here, we include additional relevant Tables and Figures.

\begin{figure*}[ht]
\centering

\begin{subfigure}[t]{.44\textwidth}
\centering
\begin{overpic}[width=\linewidth]{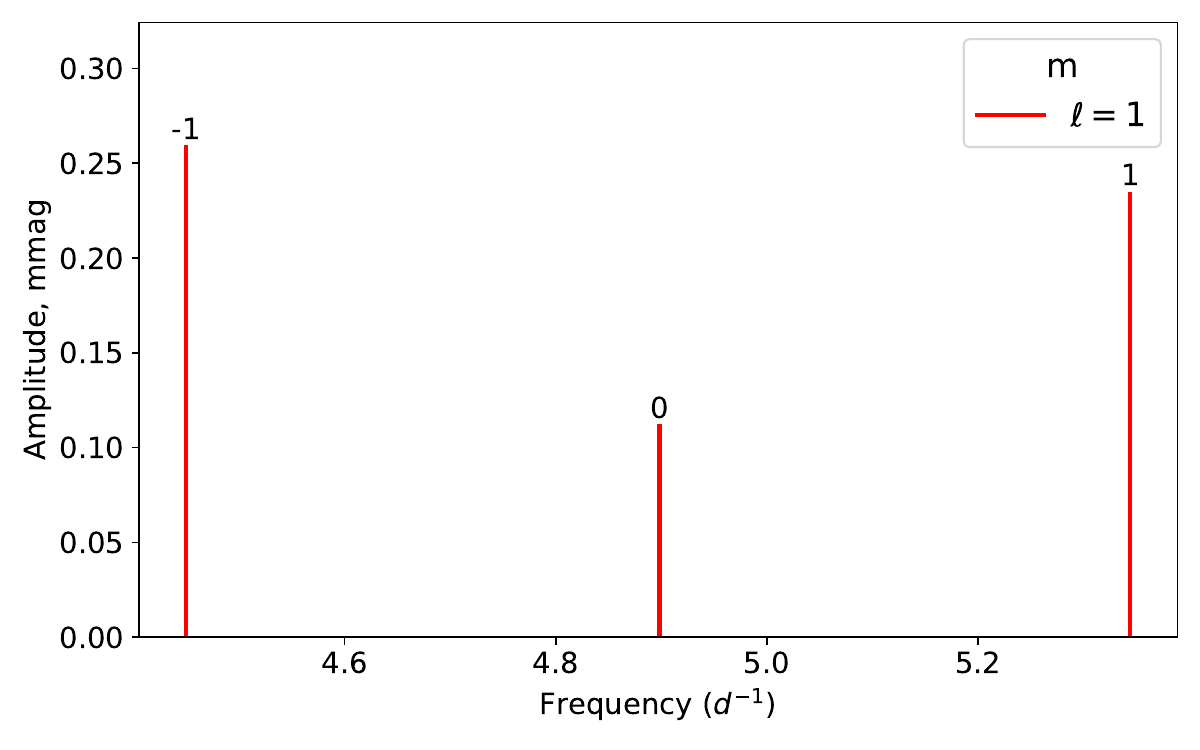}
  \put(15, 56){\small\bfseries TIC 28699677}
\end{overpic}
\caption{}\label{fig:cpd45}
\end{subfigure}
\begin{subfigure}[t]{.43\textwidth}
\centering
\begin{overpic}[width=\linewidth]{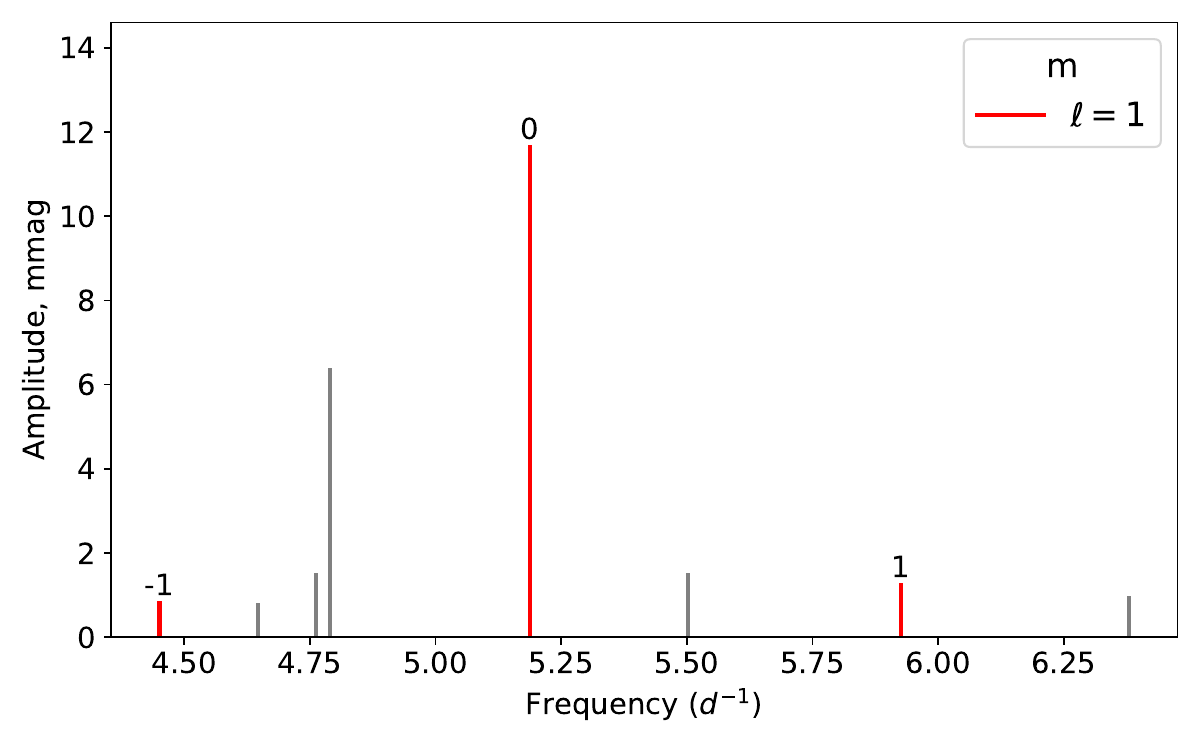}
  \put(15, 56){\small\bfseries TIC 28957011}
\end{overpic}
\caption{}\label{fig:hd101}
\end{subfigure}

\medskip

\begin{subfigure}[t]{.43\textwidth}
\centering
\begin{overpic}[width=\linewidth]{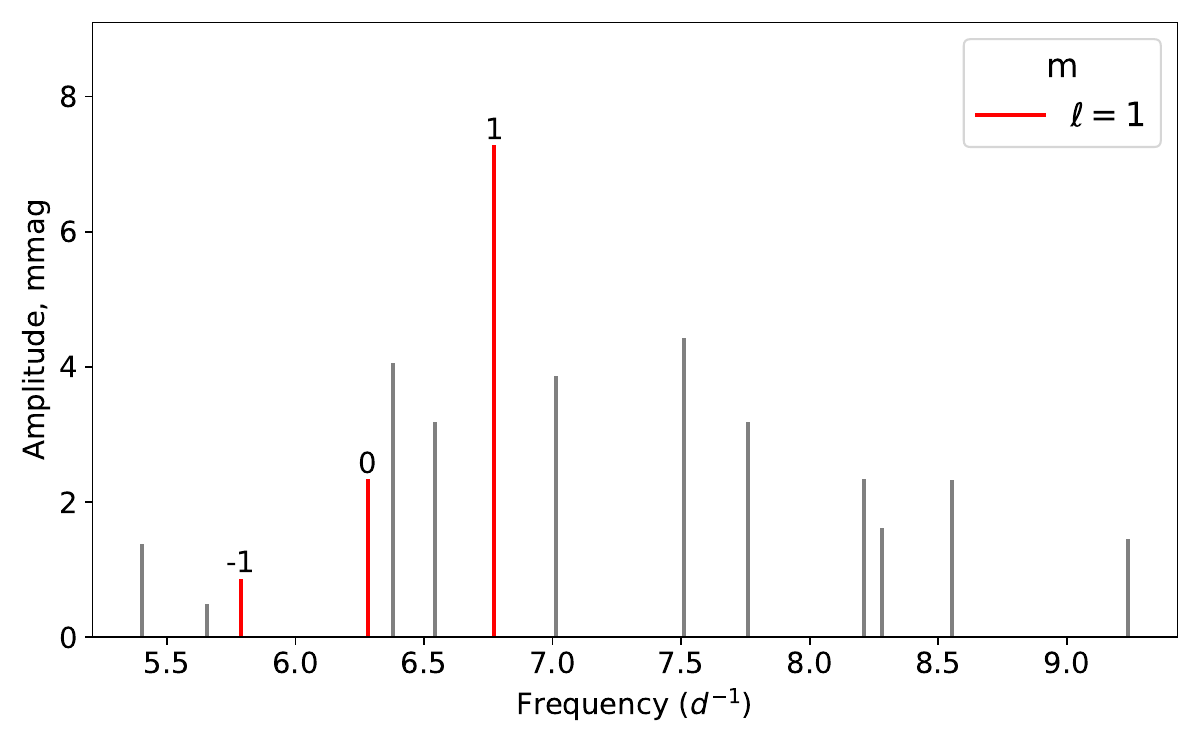}
  \put(15, 56){\small\bfseries TIC 60433558}
\end{overpic}
\caption{}\label{fig:hd108}
\end{subfigure}
\begin{subfigure}[t]{.43\textwidth}
\centering
\begin{overpic}[width=\linewidth]{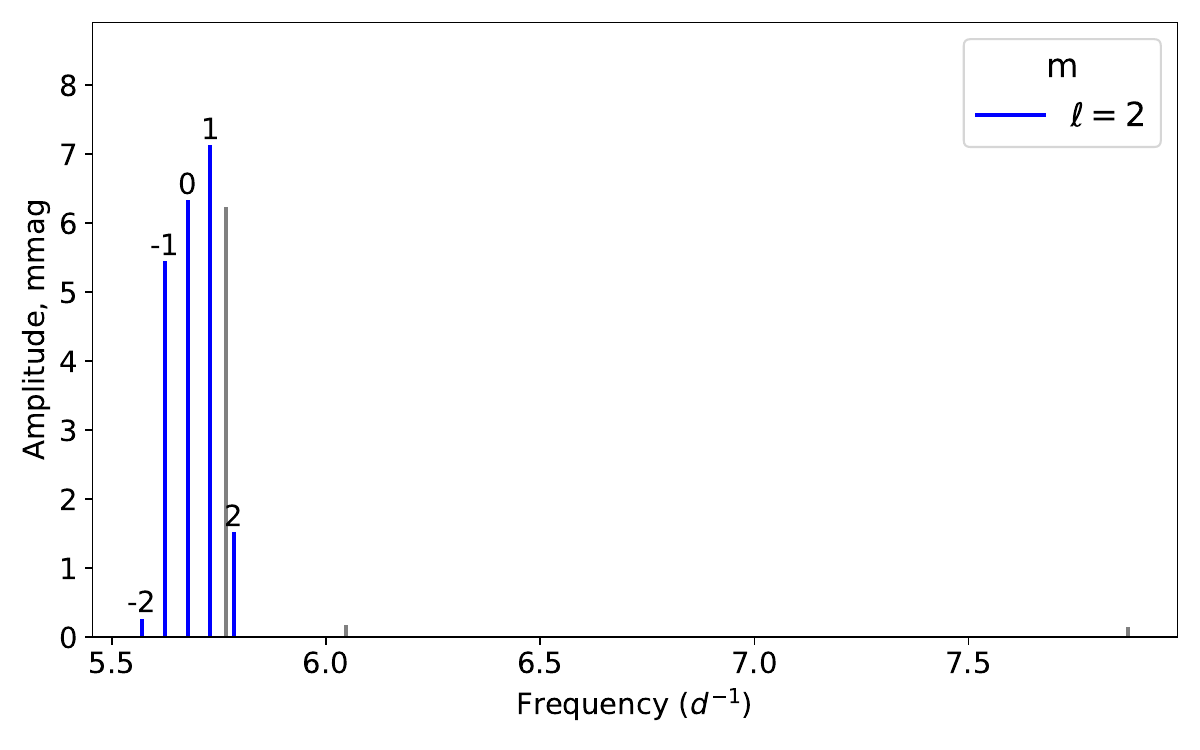}
  \put(15, 56){\small\bfseries TIC 134522557}
\end{overpic}
\caption{}\label{fig:hd1124}
\end{subfigure}

\medskip

\begin{subfigure}[t]{.43\textwidth}
\centering
\begin{overpic}[width=\linewidth]{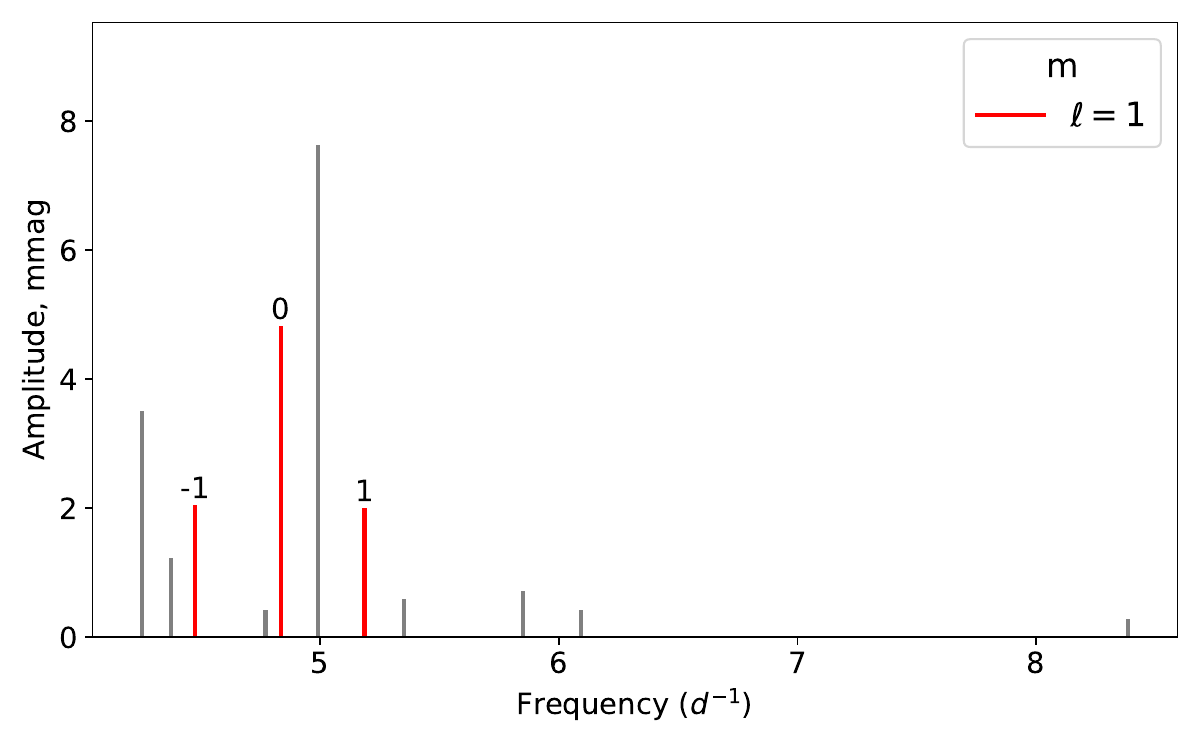}
  \put(15, 56){\small\bfseries TIC 269228628}
\end{overpic}
\caption{}\label{fig:hd157}
\end{subfigure}
\begin{subfigure}[t]{.43\textwidth}
\centering
\begin{overpic}[width=\linewidth]{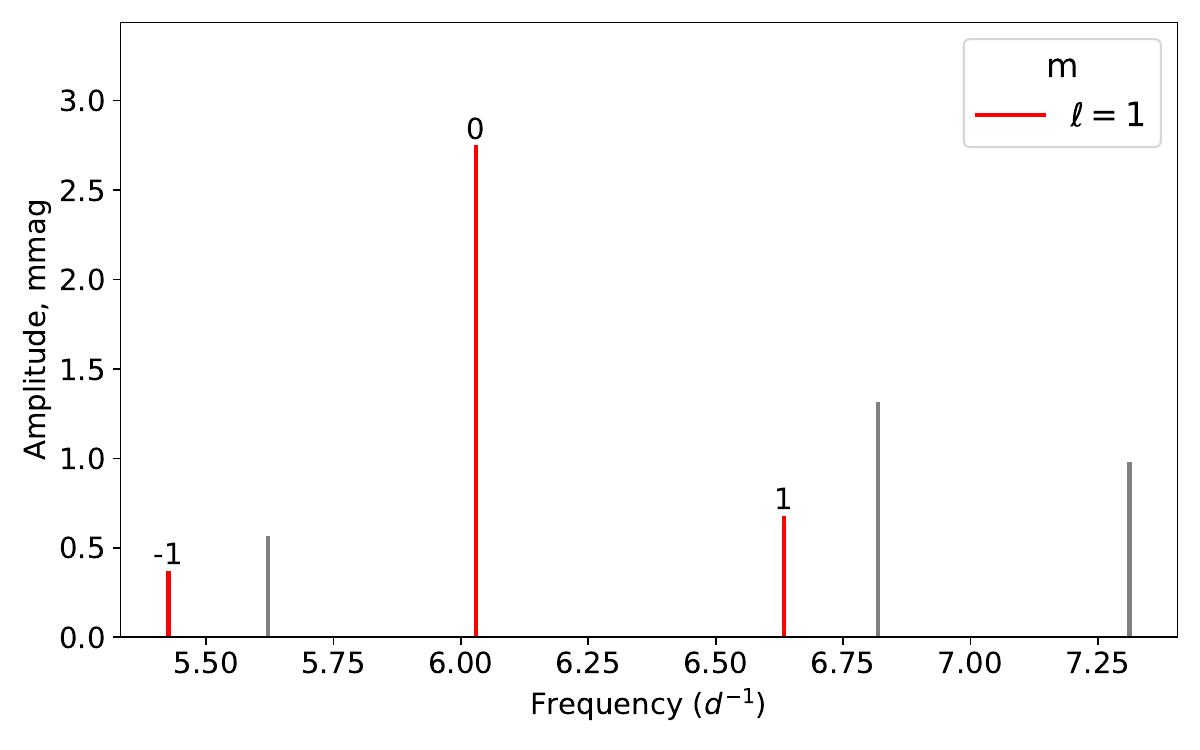}
  \put(15, 56){\small\bfseries TIC 304803692}
\end{overpic}
\caption{}\label{fig:hd1124}
\end{subfigure}

\medskip

\begin{subfigure}[t]{.43\textwidth}
\centering
\begin{overpic}[width=\linewidth]{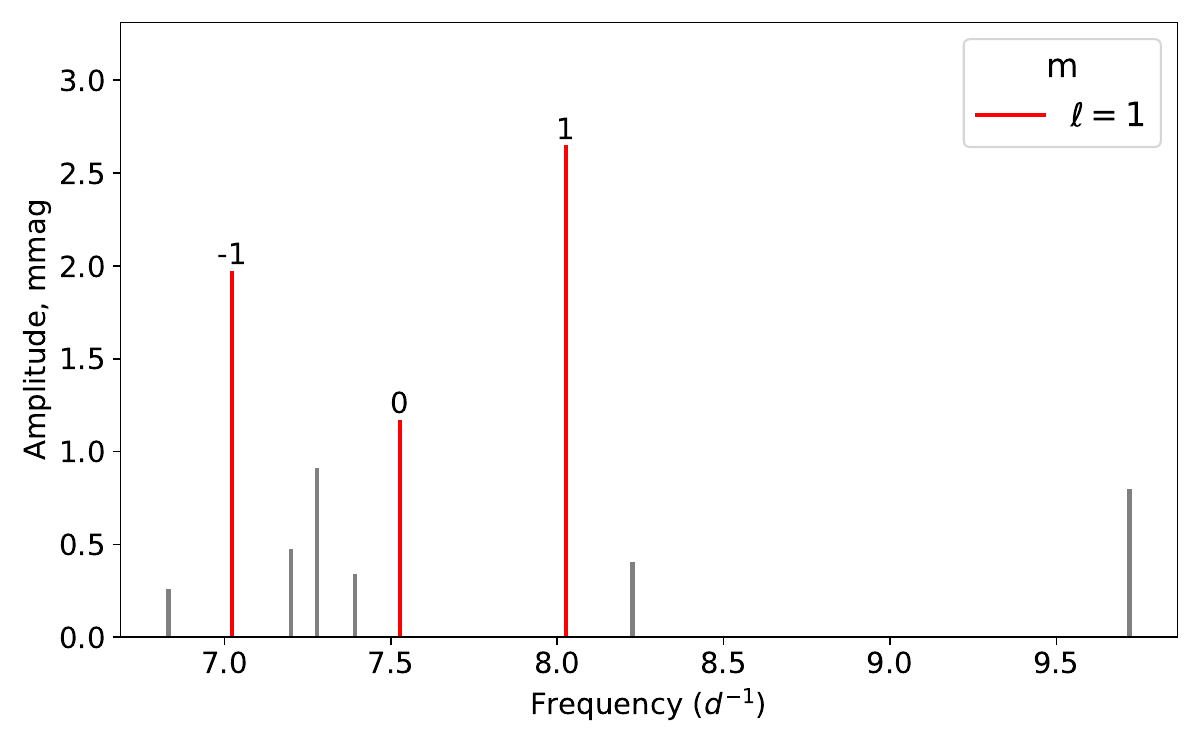}
  \put(15, 56){\small\bfseries TIC 330081196}
\end{overpic}
\caption{}\label{fig:hd157}
\end{subfigure}
\begin{subfigure}[t]{.43\textwidth}
\centering
\begin{overpic}[width=\linewidth]{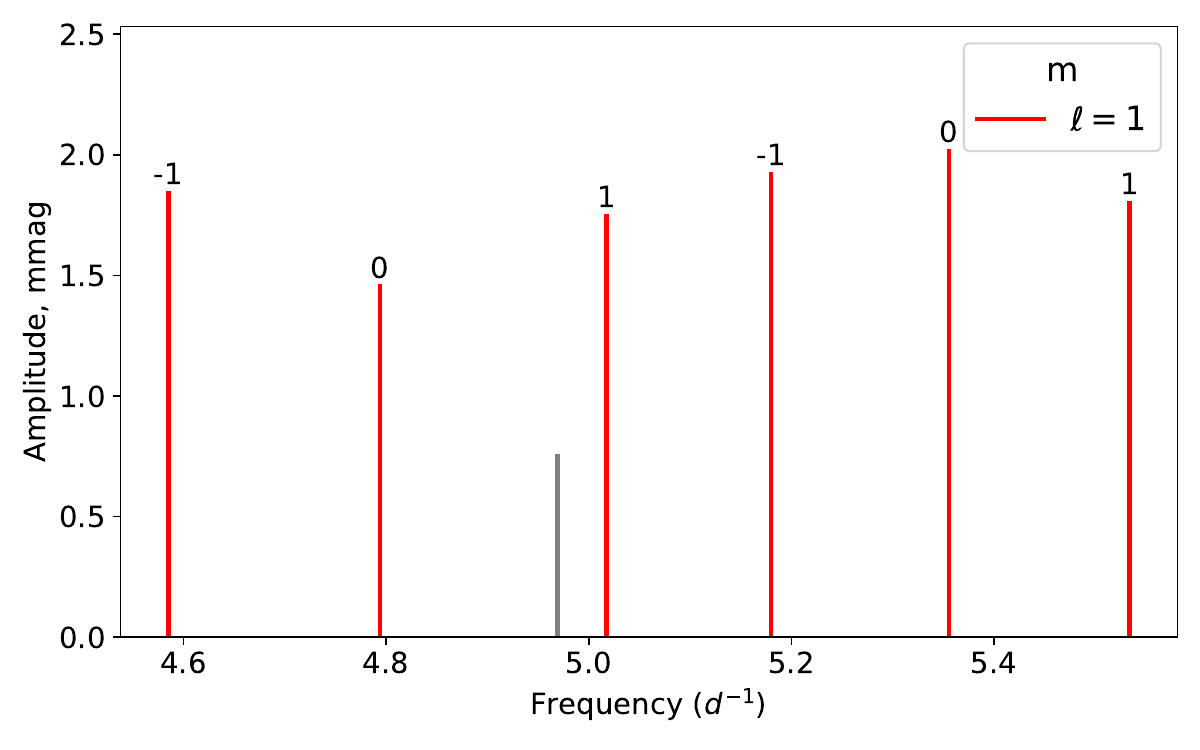}
  \put(15, 56){\small\bfseries TIC 355656323}
\end{overpic}
\caption{}\label{fig:hd254}
\end{subfigure}
\end{figure*}

\begin{figure*}[t]
    \ContinuedFloat
    \centering
\begin{subfigure}[t]{.43\textwidth}
\centering
\begin{overpic}[width=\linewidth]{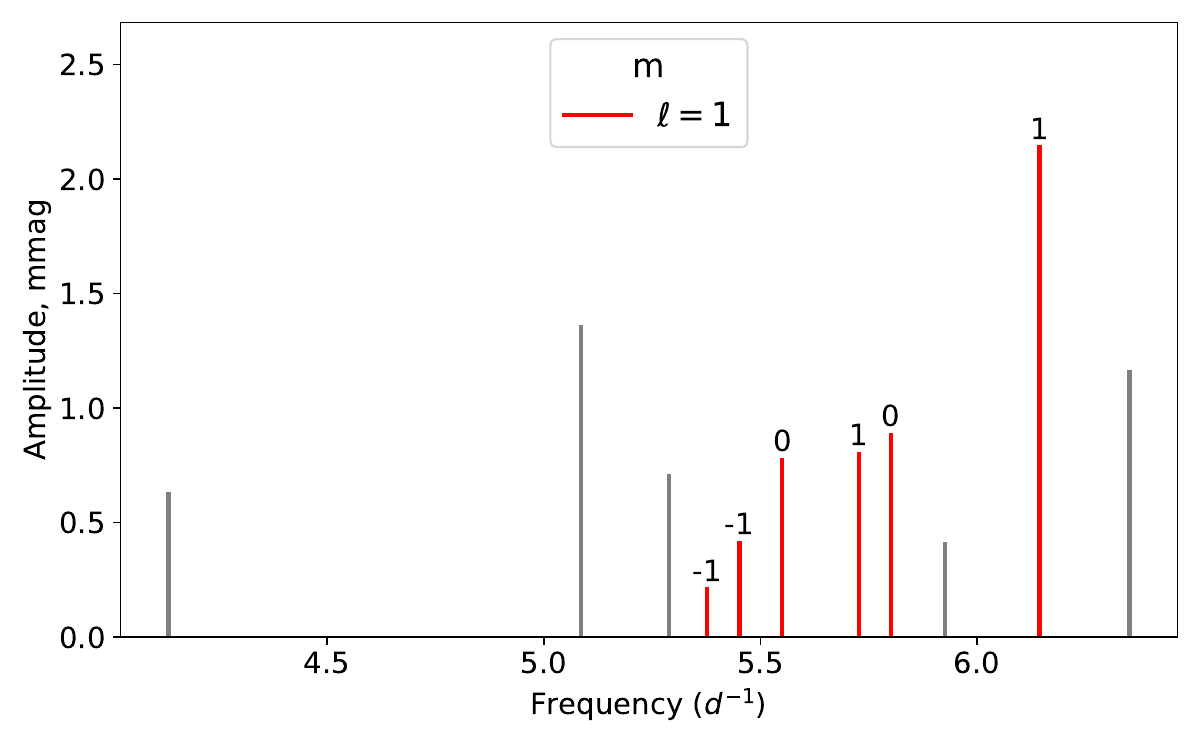}
  \put(15, 56){\small\bfseries TIC 379012185}
\end{overpic}
\caption{}\label{fig:hd329}
\end{subfigure}
\begin{subfigure}[t]{.43\textwidth}
\centering
\begin{overpic}[width=\linewidth]{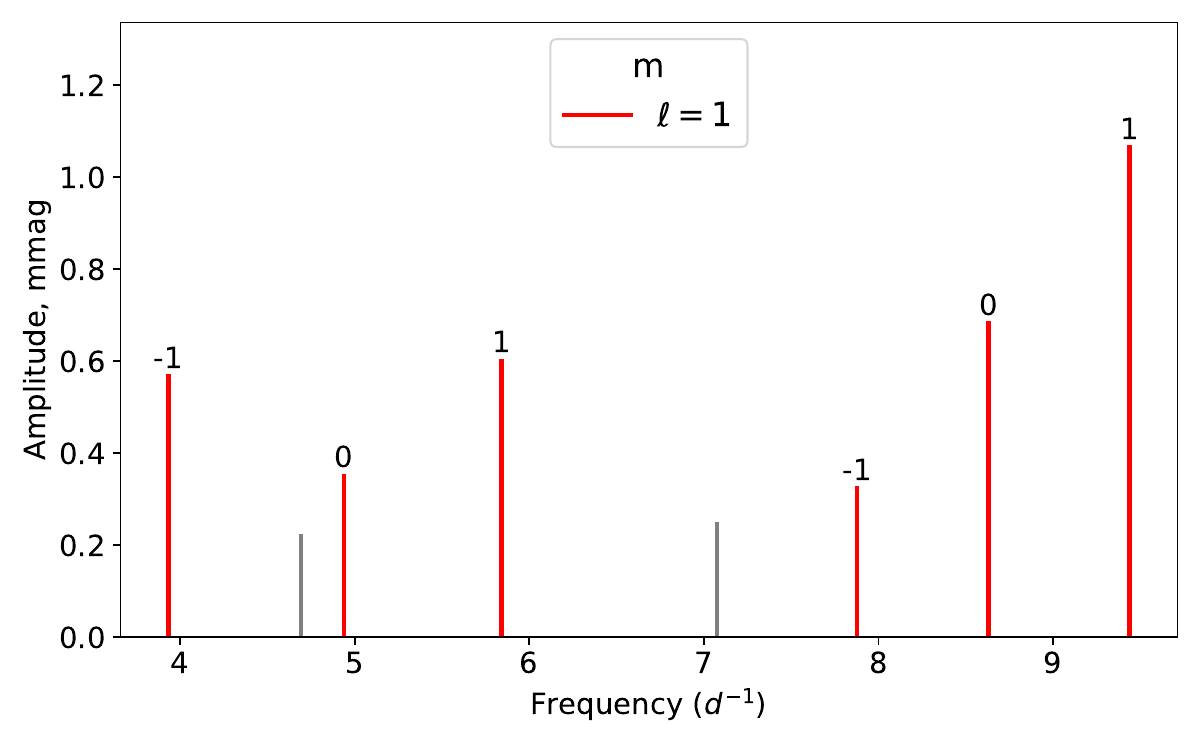}
  \put(15, 56){\small\bfseries TIC 437617380}
\end{overpic}
\caption{}\label{fig:hd339}
\end{subfigure}

\medskip

\begin{subfigure}[t]{.43\textwidth}
\centering
\begin{overpic}[width=\linewidth]{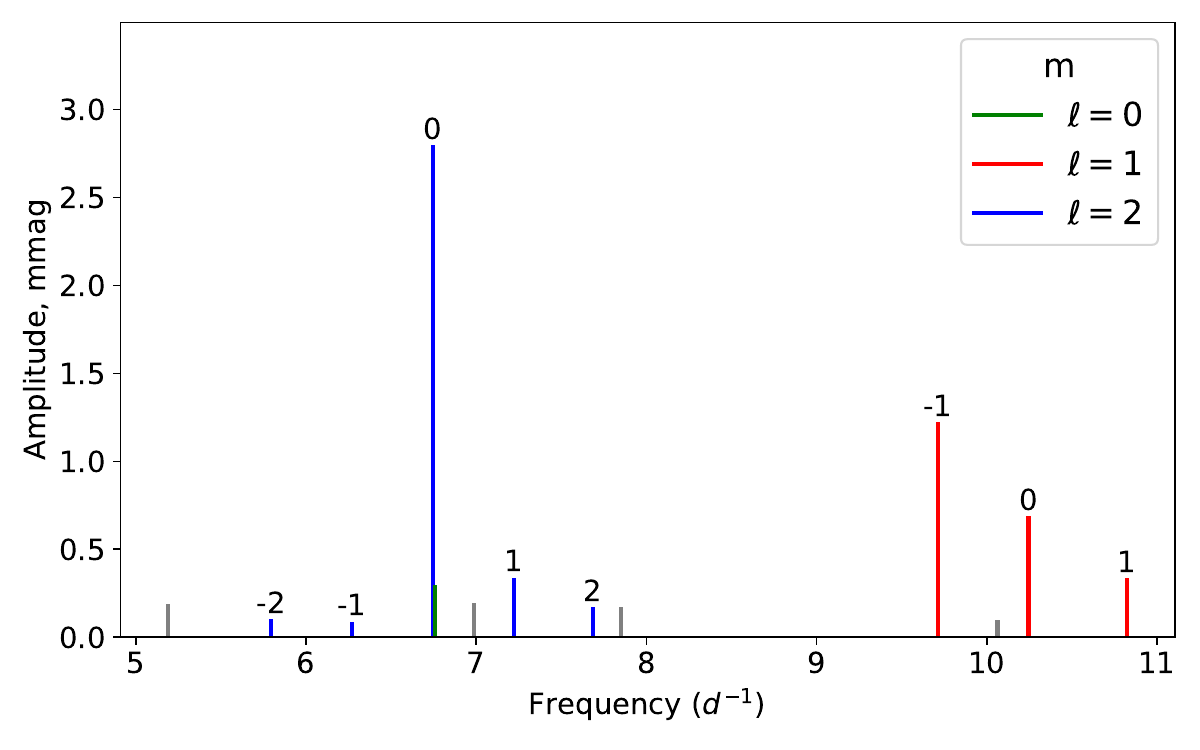}
  \put(15, 56){\small\bfseries TIC 450918869}
\end{overpic}
\caption{}\label{fig:hd927}
\end{subfigure}
\begin{subfigure}[t]{.43\textwidth}
\centering
\begin{overpic}[width=\linewidth]{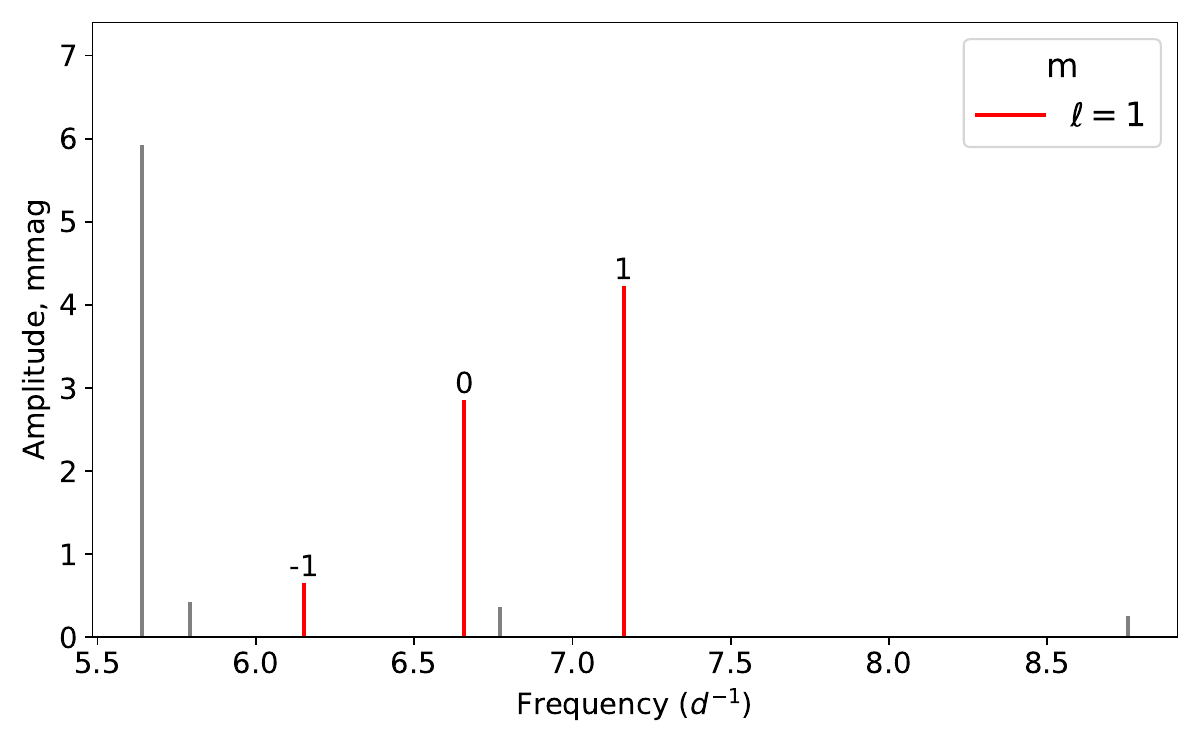}
  \put(15, 56){\small\bfseries TIC 458076434}
\end{overpic}
        \caption{}\label{fig:v1166}
\end{subfigure}

\medskip

\begin{subfigure}[t]{.43\textwidth}
\centering
\begin{overpic}[width=\linewidth]{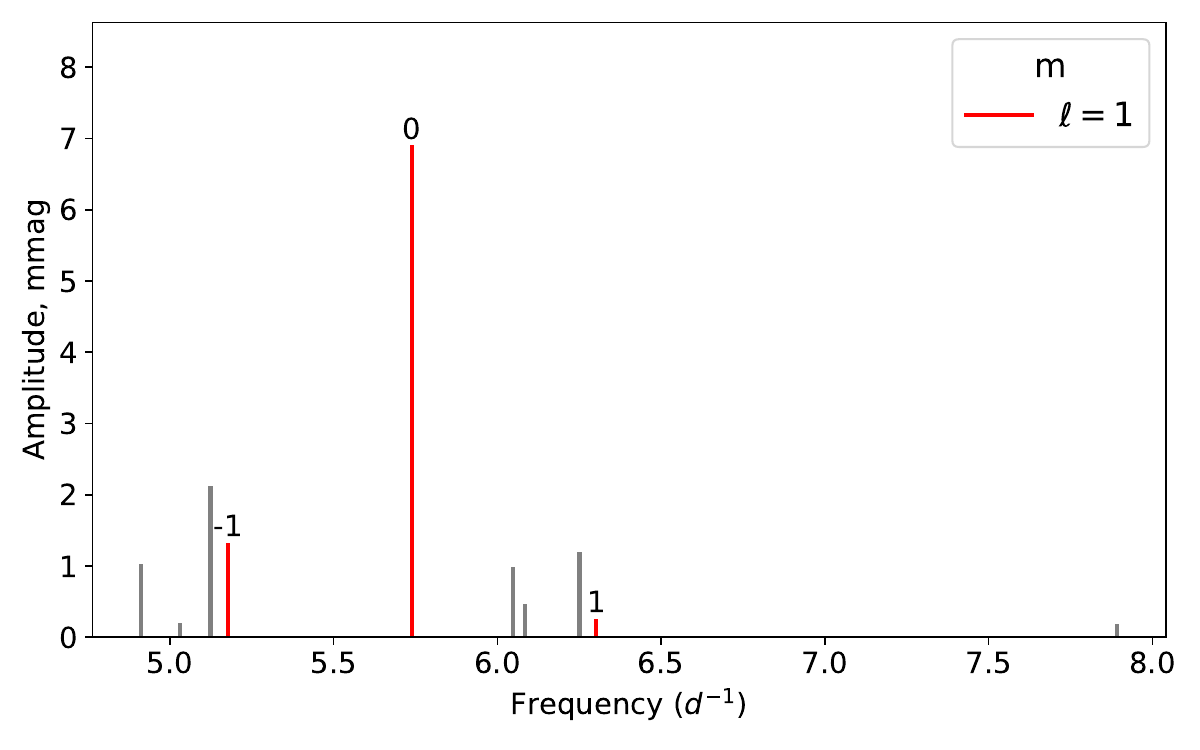}
  \put(15, 56){\small\bfseries TIC 458263480}
\end{overpic}
\caption{}\label{fig:v4386}
\end{subfigure}
\begin{subfigure}[t]{.43\textwidth}
\centering
\begin{overpic}[width=\linewidth]{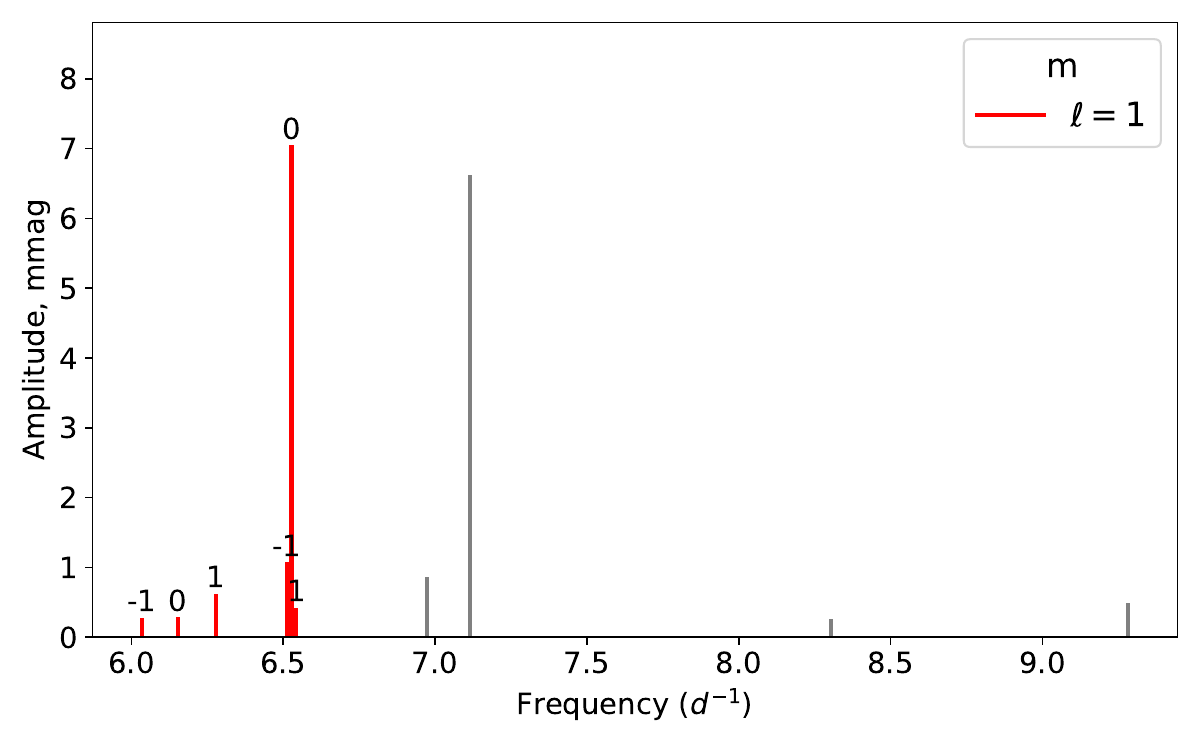}
  \put(15, 56){\small\bfseries TIC 458599043}
\end{overpic}
\caption{}\label{fig:v4386}
\end{subfigure}

\caption{Systems exhibiting rotational multiplets. Collectively, the identified pulsation modes span angular degrees $\ell = 0$ -- $2$. }\label{Appendix:Figa}
\end{figure*}

\newpage
\begin{longtable}{|*9{c|}}
\caption{Independent $\beta$ Cep pulsation frequencies and amplitudes of the systems showing rotational multiplets, where $\ell$ is the spherical degree and $m$ is the azimuthal order. Spacing is $\Delta f_{ij}$ and $\Delta f_{jk}$. Accepted significant independent $\beta$ Cep pulsations used for the analysis in this work is shown via the link \href{https://zenodo.org/records/21026255?preview=1&token=eyJhbGciOiJIUzUxMiJ9.eyJpZCI6ImEwNmY0MzAwLTY5MTEtNDIwMy1iNzY0LTM0Zjg5MjhkZDExNyIsImRhdGEiOnt9LCJyYW5kb20iOiJlYjc3ZTFlY2NjNmNjYzVjMzMzZjRlZjBmMTY2OTE3NSJ9.x9dFNuG3l_XiW95fASsFIUDvVDBX_VsWQGpwp9L16qey7xb-04bnGmqwxU5brwcM8wr9SxxKTUsOJe_LLO6kXg}{\it{list of pulsations}}. The first frequency in the online table represents the dominant pulsation in each target}\label{appendix:tab-rot-multiplet} \\

\hline
\hline
TIC ID & $P$ (d) & $f$ (d$^{-1}$) & A (mmag) & $\ell$ & m & spacing  (d$^{-1}$) & [$A_{s}$  ($d^{-1}$); $A_{\mathrm{norm}}$] & $f_{\rm rot}$ (d$^{-1}$) \\
\hline
\endfirsthead

\hline
\hline
TIC ID & $P$ (d) & $f$ (d$^{-1}$) & A (mmag) & $\ell$ & m & spacing \, d$^{-1}$ & [$A_{s}\, d^{-1}$; $A_{\mathrm{norm}}$] & $f_{\rm rot}$ (d$^{-1}$) \\
\hline
\endhead

\hline
\multicolumn{9}{r}{\textit{Continued on next page}} \\
\hline
\endfoot

\hline
\endlastfoot

28699677	&22.9444(5) & & & & & & & \\
&    &	4.44995(3)	&	0.26(1)	&	1	&	$-1$	&	\multirow{2}{*}{0.4483}	&		&		\\
&	&	4.89821(7)	&	0.11(1)	&	1	&	0	&	\multirow{2}{*}{0.4457}	&	[0.0025; 0.003]	&	0.44698(5)	\\
&	&	5.34393(3)	&	0.24(1)	&	1	&	1	&		&		&		\\
    \hline
28957011	& 2.7139(1)& & & & & & & \\
&   &	4.451(2)	&	0.9(1)	&	1	&	$-1$	&	\multirow{2}{*}{0.7364}	&		&		\\
&	&	5.1873(1)	&	11.7(1)	&	1	&	0	&	\multirow{2}{*}{0.7379}	&	[0.0016; 0.001]	&	0.737(1)	\\
&	&	5.925(1)	&	1.3(1)	&	1	&	1	&	&		&		\\
&	&	5.5018(9)	&	1.5(1)	&		&		&		&		&		\\
&	&	4.7621(9)	&	1.5(1)	&		&		&		&		&		\\
&	&	6.379(2)	&	0.9(1)	&		&		&		&		&		\\
&	&	4.646(2)	&	0.8(1)	&		&		&		&		&		\\
&	&	4.79005(2)	&	6.4(1)	&		&		&		&		&		\\
    \hline
60433558	& 10.798(1)& & & & & & &\\
& &	5.786807(6)	&	0.87(1)	&	1	&	$-1$	&	\multirow{2}{*}{0.4941}	&		&		\\
&	&	6.280927(2)	&	2.33(1)	&	1	&	0	&	\multirow{2}{*}{0.4924}	&	[0.0018; 0.002] &	0.493238(3)	\\
&	&	6.7732821(7)	&	7.28(1)	&	1	&	1	&		&		&		\\
&	&	7.512030(1)	&	4.43(1)	&		&		&		&		&		\\
&	&	6.377774(1)	&	4.06(1)	&		&		&		&		&		\\
&	&	7.014646(1)	&	3.87(1)	&		&		&		&		&		\\
&	&	7.759924(2)	&	3.19(1)	&		&		&		&		&		\\
&	&	6.542596(2)	&	3.18(1)	&		&		&		&		&		\\
&	&	8.212493(2)	&	2.34(1)	&		&		&		&		&		\\
&	&	8.554092(2)	&	2.32(1)	&		&		&		&		&		\\
&	&	8.282639(3)	&	1.62(1)	&		&		&		&		&		\\
&	&	9.238721(4)	&	1.45(1)	&		&		&		&		&		\\
&	&	5.401952(4)	&	1.38(1)	&		&		&		&		&		\\
&	&	5.65661(1)	&	0.49(1)	&		&		&		&		&		\\
   \hline
134522557	& 5.730(2)& & & & & & &\\
& &	5.57001(1)	&	0.27(2)	&	2	&	$-2$	&	\multirow{2}{*}{0.1079}	&		&		\\
&	&	5.6237481(7)	&	5.45(2)	&	2	&	$-1$	&	\multirow{2}{*}{0.0542}&		&		\\
&	&	5.6779561(6)	&	6.33(2)	&	2	&	0	&	\multirow{2}{*}{0.0521}	&	[0.0021, 0.0005; &	0.053844(3)	\\
&	&	5.7300571(5)	&	7.13(2)	&	2	&	1	&	\multirow{2}{*}{0.1074}	&	0.02, 0.002]	&		\\
&	&	5.785393(3)	&	1.52(2)	&	2	&	2	&	&		&		\\
&	&	5.7657150(6)	&	6.24(2)	&		&		&		&		&		\\
&	&	6.04692(2)	&	0.18(2)	&		&		&		&		&		\\
&	&	7.87248(3)	&	0.14(2)	&		&		&		&		&		\\
    \hline
269228628	&2.422(1) & & & & & &&\\
& &	4.477037(4)	&	2.05(2)	&	1	&	$-1$	&	\multirow{2}{*}{0.3579}&		&		\\
&	&	4.835017(1)	&	4.83(2)	&	1	&	0	&	\multirow{2}{*}{0.3512}	&	[0.0068; 0.009]	&	0.354582(3)	\\
&	&	5.186201(4)	&	2.00(2)	&	1	&	1	&		&		&		\\
&	&	4.9918992(9)	&	7.62(2)	&		&		&		&		&		\\
&	&	4.252738(2)	&	3.51(2)	&		&		&		&		&		\\
&	&	4.376108(6)	&	1.23(2)	&		&		&		&		&		\\
&	&	5.84860(1)	&	0.72(2)	&		&		&		&		&		\\
&	&	5.35212(1)	&	0.59(2)	&		&		&		&		&		\\
&	&	4.77074(2)	&	0.43(2)	&		&		&		&		&		\\
&	&	6.09405(2)	&	0.42(2)	&		&		&		&		&		\\
&	&	8.38589(3)	&	0.29(2)	&		&		&		&		&		\\
    \hline
304803692	& 4.92870(2)& & & & & & &\\
& &	5.426(2)	&	0.37(3)	&	1	&	$-1$	&	\multirow{2}{*}{0.6039}	&		&		\\
&	&	6.0300(2)	&	2.75(3)	&	1	&	0	&	\multirow{2}{*}{0.6046}	&	[0.0006; 0.0005]	&	0.604(1) \\
&	&	6.6346(9)	&	0.68(3)	&	1	&	1	&		&		&		\\
&	&	6.8180(6)	&	1.31(3)	&		&		&		&		&		\\
&	&	7.3121(5)	&	0.98(3)	&		&		&		&		&		\\
&	&	5.622(1)	&	0.57(3)	&		&		&		&		&		\\
    \hline
330081196	& 4.80323(6)& & & & & & &\\
& &	6.83178(5)	&	0.26(2)	&	1	&	$-1$	&	&		&		\\
&	&	7.022813(6)	&	1.98(2)	&	1	&	$-1$	&\multirow{2}{*}{0.5048	}	&		&		\\
&	&	7.52762(1)	&	1.17(2)	&	1	&	0	&	\multirow{2}{*}{0.4987}	&[0.006; 0.006]	&0.59951(3)	\\
&	&	8.026362(5)	&	2.65(2)	&	1	&	1	&	&		&		\\

&	&	8.22630(3)	&	0.41(2)	&	1	&	1	&		&		&		\\
&	&	9.71971(2)	&	0.79(2)	&		&		&		&		&		\\
&	&	7.19968(3)	&	0.48(2)	&		&		&		&		&		\\
&	&	7.39349(4)	&	0.34(2)	&		&		&		&		&		\\
&	&	7.27848(1)	&	0.91(2)	&		&		&		&		&		\\
    \hline
355656323	& 5.1929(1)& & & & & & &\\
& &	5.18011(1)	&	1.93(4)	&	1	&	$-1$	&	\multirow{2}{*}{0.1753}	&		&		\\
&	&	5.35544(1)	&	2.02(4)	&	1	&	0	&	\multirow{2}{*}{0.1781}	&	[0.0027; 0.008]	&	0.17669(1)	\\
&	&	5.53351(1)	&	1.81(4)	&	1	&	1	&		&		&		\\
&	&	4.58521(1)	&	1.85(4)	&	1	&	$-1$	& \multirow{2}{*}{0.3838}		&		&		\\
&	&	4.79358(2)	&	1.46(4)	&	1	&	0	&	\multirow{2}{*}{0.3864}	&	[0.0026; 0.0034] &	0.38510(1)	\\
&	&	5.01749(1)	&	1.75(4)	&	1	&		1&		&		&		\\
&	&	4.96904(3)	&	0.76(4)	&		&		&		&		&		\\
    \hline
379012185	&4.7447(1) & & & & & & & \\
& &	5.4513(3)	&	0.422(7)	&	1	&	$-1$	&	\multirow{2}{*}{0.3503}	&		&		\\
&	&	5.8017(2)	&	0.892(7)	&	1	&	0	&	\multirow{2}{*}{0.3420}	&	[0.0083; 0.01] &	\\
&	&	6.14371(7)	&	2.146(7)	&	1	&	1	&		&		& 0.1751(4)\\
&	&	5.3767(6)	&	0.217(7)	&	1	&	$-1$	&	\multirow{2}{*}{0.1737}	&		&		\\
&	&	5.5505(2)	&	0.783(7)	&	1	&	0	&	\multirow{2}{*}{0.1763}	&	[0.0026; 0.007]	&		\\
&	&	5.7268(2)	&	0.810(7)	&	1	&	1	&	&		&		\\
&	&	5.0863(1)	&	1.362(7)	&		&		&		&		&		\\
&	&	6.3514(1)	&	1.168(7)	&		&		&		&		&		\\
&	&	5.2891(2)	&	0.713(7)	&		&		&		&		&		\\
&	&	4.1337(2)	&	0.635(7)	&		&		&		&		&		\\
&	&	5.9259(3)	&	0.416(7)	&		&		&		&		&		\\
    \hline
437617380$^{*}$	&5.372(1) & & & & & & &\\
& &	7.8788(6)	&	0.33(2)	&	1	&	$-1$	&	\multirow{2}{*}{0.7516}	&		&		\\
&	&	8.6304(3)	&	0.69(2)	&	1	&	0	&	\multirow{2}{*}{0.8073}	&	[0.0557; 0.04] &		\\
&	&	9.4377(2)	&	1.07(2)	&	1	&	1	&		&		&0.8666(7)	\\
&	&	3.9334(3)	&	0.57(2)	&	1&	$-1$	&	\multirow{2}{*}{1.0062}	&		&		\\
&	&	4.9396(7)	&	0.36(2)	&	1	&	0	&	\multirow{2}{*}{0.9012}	&	[0.1; 0.05]	&		\\
&	&	5.8408(3)	&	0.61(2)	&	1	&	1	&	&		&		\\
& &7.0730(8) &0.25(2) & & & & & \\
&	&	4.6935(8)	&	0.22(2)	&		&		&		&		&		\\
    \hline
450918869	& 4.222(5)& & & & & & &\\
& &	6.75639(2)	&	0.29(1)	&	0	&		&		&		&		\\
&	&	5.79438(5)	&	0.10(1)	&	2	&	$-2$	&	\multirow{2}{*}{0.9495}	&		&		\\
&	&	6.26911(5)	&	0.09(1)	&	2	&	$-1$	&	\multirow{2}{*}{0.4748}	&		&		\\
&	&	6.743891(2)	&	2.79(1)	&	2	&	0	&	\multirow{2}{*}{0.4778}	&	[0.003, 0.008; 	&	0.47275(1)\\
&	&	7.22166(1)	&	0.34(1)	&	2	&	1	&		\multirow{2}{*}{0.9415}&	0.003, 0.004]	&		\\
&	&	7.68537(3)	&	0.17(1)	&	2	&	2	&		&		&		\\
&	&	9.711631(4)	&	1.22(1)	&		1&	$-1$	&	\multirow{2}{*}{0.5340}	&		&		\\
&	&	10.245958(7)	&	0.69(1)	&	1	&	0	&	\multirow{2}{*}{0.5780}	&	0.044	&	0.556(2)	\\
&	&	10.82348(1)	&	0.34(1)	&	1	&	1	&	&		&		\\
&	&	6.98505(2)	&	0.19(1)	&		&		&		&		&		\\
&	&	7.84854(3)	&	0.17(1)	&		&		&		&		&		\\
&	&	10.06382(5)	&	0.10(1)	&		&		&		&		&		\\
&	&	5.19053(2)	&	0.19(1)	&		&		&		&		&		\\
    \hline
458076434	& 8.3249(8)& & & & & & &\\
&   &	6.15265(3)	&	0.65(3)	&	1	&	$-1$	&	\multirow{2}{*}{0.5062}	&		&		\\
&	&	6.658866(7)	&	2.86(3)	&	1	&	0	&		\multirow{2}{*}{0.5032}&	[0.003; 0.003] &	0.50470(2)	\\
&	&	7.162049(5)	&	4.23(3)	&	1	&	1	&	&		&		\\
&	&	5.640013(3)	&	5.92(3)	&		&		&		&		&		\\
&	&	5.79341(5)	&	0.43(3)	&		&		&		&		&		\\
&	&	6.77019(6)	&	0.36(3)	&		&		&		&		&		\\
&	&	8.75539(8)	&	0.26(3)	&		&		&		&		&		\\
    \hline
458263480	&4.618(1) & & & & & & &\\
& &	5.17808(1)	&	1.33(2)	&	1	&	$-1$	&	\multirow{2}{*}{0.5613}	&		&		\\
&	&	5.739375(2)	&	6.90(2)	&	1	&	0	&	\multirow{2}{*}{0.5626}	&	[0.0013; 0.001] &	0.56195(3)	\\
&	&	6.30198(6)	&	0.26(2)	&	1	&	1	&	&		&		\\
&	&	5.124229(9)	&	2.12(2)	&		&		&		&		&		\\
&	&	6.25069(1)	&	1.19(2)	&		&		&		&		&		\\
&	&	4.91264(2)	&	1.03(2)	&		&		&		&		&		\\
&	&	6.04805(2)	&	0.98(2)	&		&		&		&		&		\\
&	&	6.08365(4)	&	0.46(2)	&		&		&		&		&		\\
&	&	5.03004(8)	&	0.20(2)	&		&		&		&		&		\\
&	&	7.89274(9)	&	0.18(2)	&		&		&		&		&		\\
    \hline
458599043	&14.46(2) & & & & & & & \\
& &	6.51201(3)	&	1.07(5)	&	1	&	$-1$	&	\multirow{2}{*}{0.0149}	&		&		\\
&	&	6.526882(5)	&	7.04(5)	&	1	&	0	&	&	[0.0014; 0.04]	&	\\
&	&	6.54317(9)	&	0.43(5)	&	1	&	1	&	0.0163&		&	0.069(1)	\\
&	&	6.0341(1)	&	0.28(5)	&	1	&	$-1$	&	\multirow{2}{*}{0.1185}	&		&		\\
&	&	6.1525(1)	&	0.29(5)	&	1	&	0	&	\multirow{2}{*}{0.1251}	&	[0.0066; 0.03]	&	\\
&	&	6.27761(6)	&	0.62(5)	&	1	&	1	&		&		&		\\
&	&	7.116112(6)	&	6.62(5)	&		&		&		&		&		\\
&	&	6.97419(4)	&	0.86(5)	&		&		&		&		&		\\
&	&	9.28237(8)	&	0.49(5)	&		&		&		&		&		\\
&	&	8.3049(1)	&	0.27(5)	&		&		&		&		&		\\
\hline\hline    

\end{longtable}

\begin{figure*}[htbp]
\centering
\includegraphics[width=0.45\linewidth]{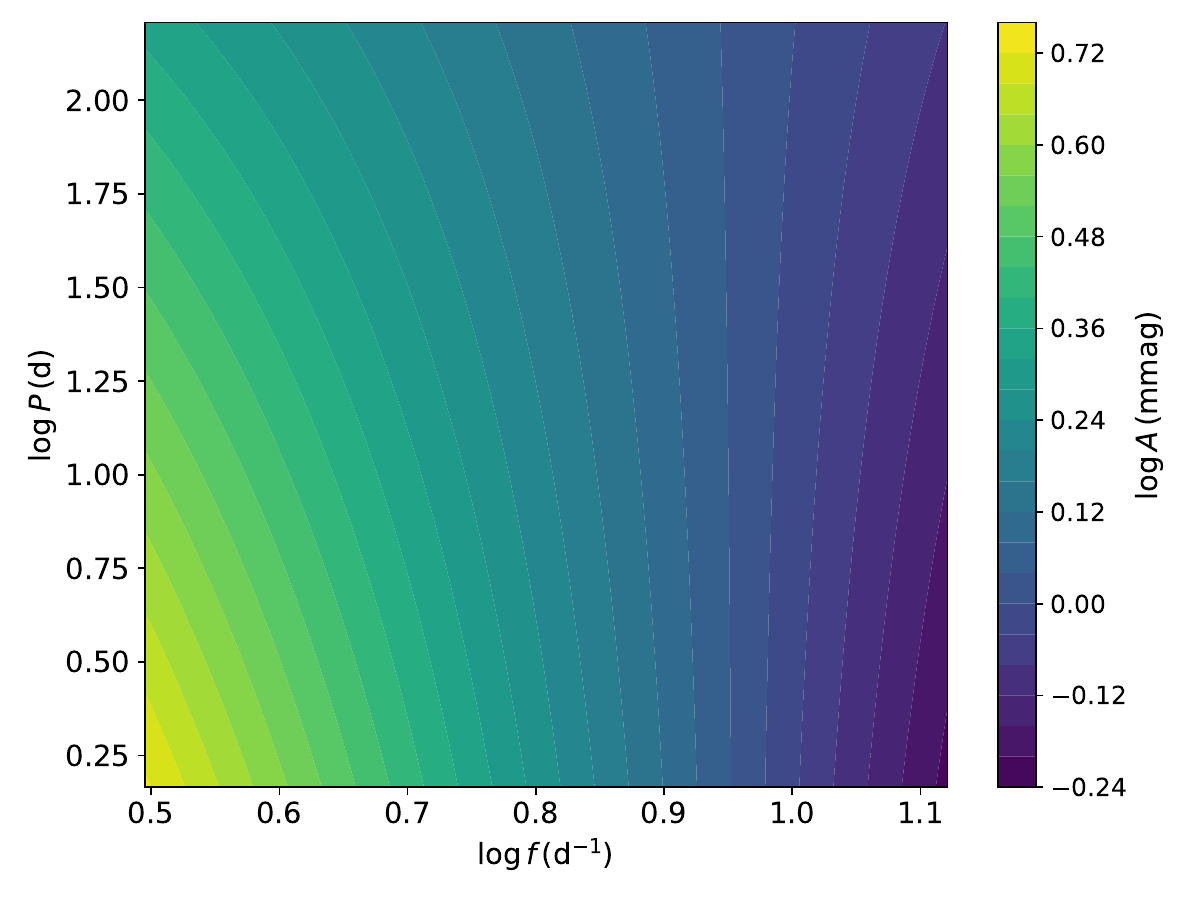}
\includegraphics[width=0.45\linewidth]{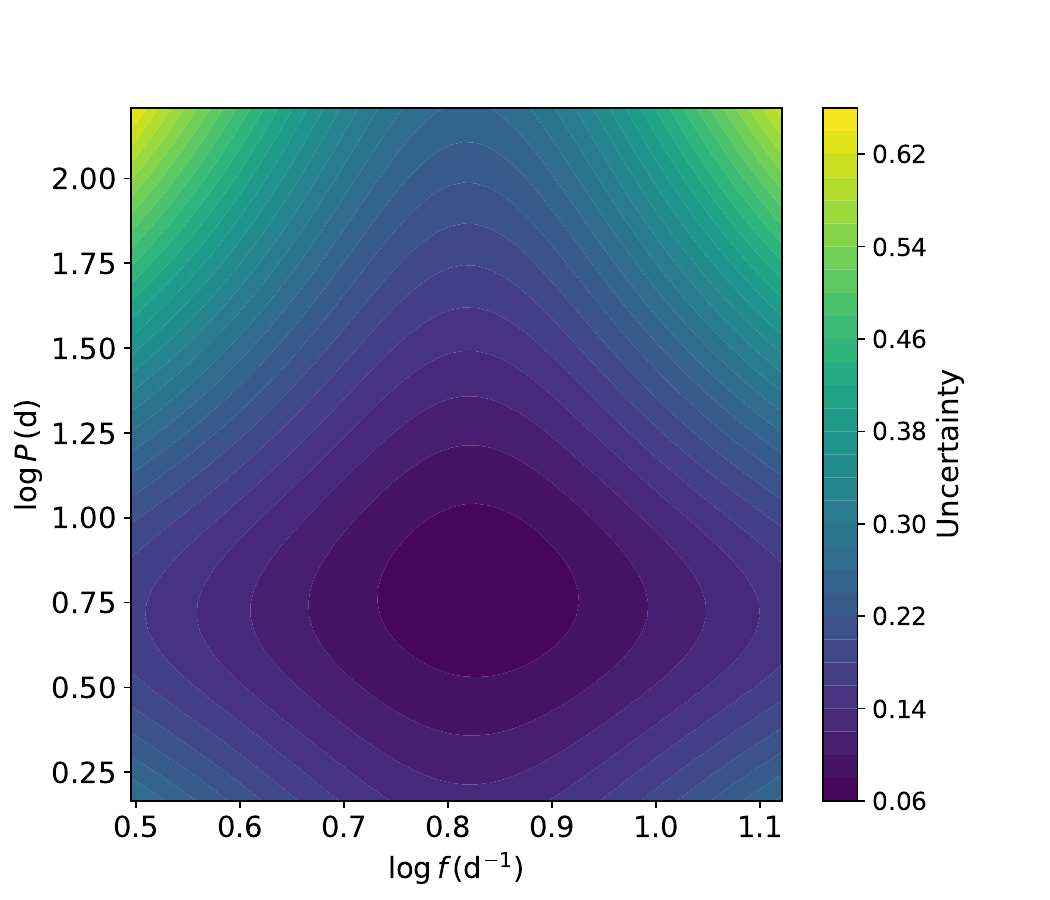}
\caption{The left panel shows the relationship among a few pulsation and binary parameters, where $P$ is the orbital period measured in days, $f$ are the pulsation frequencies, and $A$ are the pulsation amplitudes. The model uncertainty is shown in the right panel.}\label{fig:fig_model_with_interaction}

\end{figure*}

\begin{figure*}[htbp]
\centering
\includegraphics[width=0.7\linewidth]{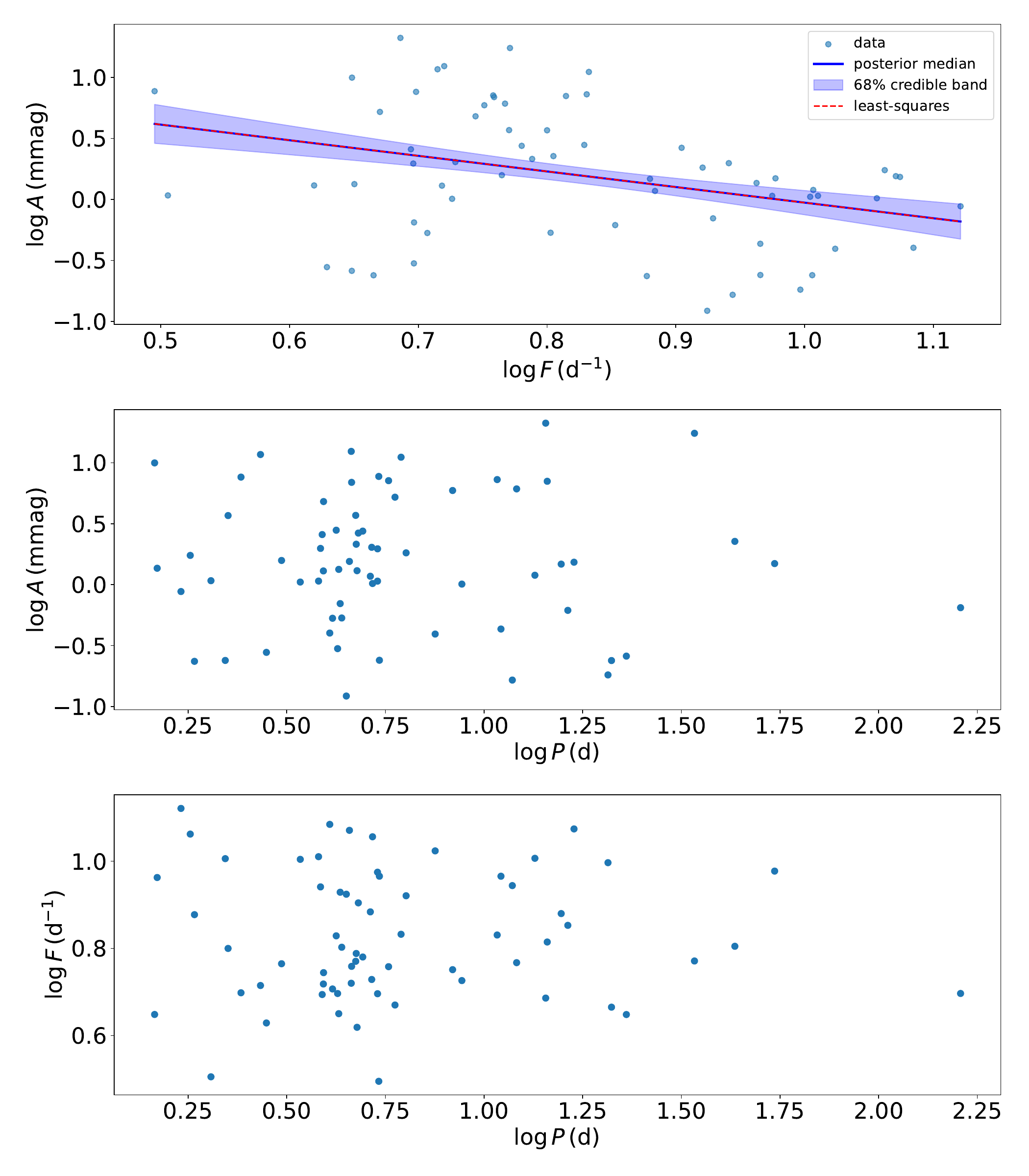}
\caption{Results of the univariate regression analysis between orbital and pulsation parameters. The ordinary least-squares fit in the top panel yielded a Pearson correlation coefficient $r=-0.35$ with  a frequentist $p$-value of $=0.005$.}\label{fig:appendix_univariate_reg}
\end{figure*}

\begin{figure*}[htbp]
\centering
\includegraphics[width=0.49\linewidth]{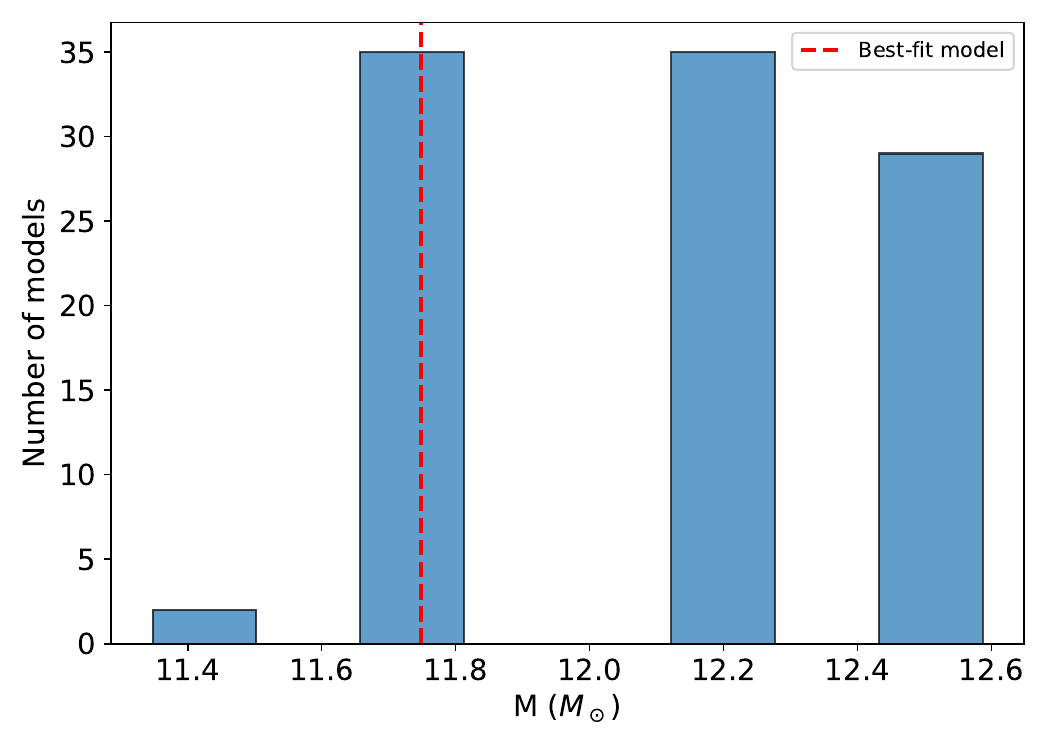}
\includegraphics[width=0.49\linewidth]{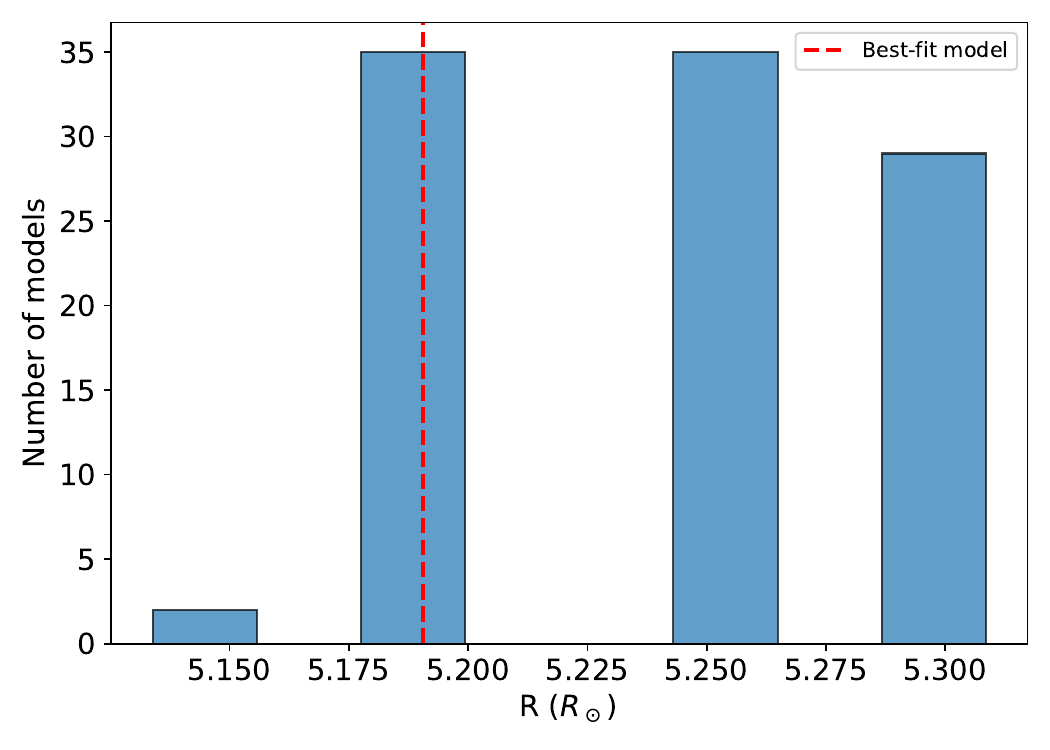}
\caption{Example plots of the unweighted distributions of the derived parameters. The distributions of mass (left panel) and radius (right panel) of HD~112485  are shown for the $\chi^2$ models compatible with the spectroscopic constraints. The gaps in the unweighted distributions reflect the discrete sampling of the stellar-model grid and the subsequent selection of only models satisfying the adopted constraints. They are therefore not interpreted as evidence for physically distinct populations. }\label{fig:appendix_unweighted_distribution}
\end{figure*}

\end{appendix}

\end{document}